\documentclass[twocolumn,pra,showpacs]{revtex4}
\pdfoutput=1
\usepackage{setspace}
\usepackage{amsfonts,amsmath,amssymb}
\usepackage{txfonts}
\usepackage{graphicx, color}
\usepackage{hyperref}
\usepackage{bm}

\begin{document}

\title{Finite-temperature dynamics of a single vortex in a Bose-Einstein condensate: 
 Equilibrium precession and rotational symmetry breaking}

\author{T. M. Wright}
\affiliation{Jack Dodd Centre for Quantum Technology, Department of Physics, University of Otago, PO Box 56, Dunedin, New Zealand}

\author{A. S. Bradley}
\affiliation{Jack Dodd Centre for Quantum Technology, Department of Physics, University of Otago, PO Box 56, Dunedin, New Zealand}

\author{R. J. Ballagh}
\affiliation{Jack Dodd Centre for Quantum Technology, Department of Physics, University of Otago, PO Box 56, Dunedin, New Zealand}

\begin{abstract}
We consider a finite-temperature Bose-Einstein condensate in a quasi-two-dimensional trap containing a single precessing vortex.  We find that such a configuration arises naturally as an ergodic equilibrium of the projected Gross-Pitaevskii equation, when constrained to a finite conserved angular momentum.  In an isotropic trapping potential the condensation of the classical field into an off-axis vortex state breaks the rotational symmetry of the system.  We present a methodology to identify the condensate and the Goldstone mode associated with the broken rotational symmetry in the classical-field model.  We also examine the variation in vortex trajectories and thermodynamic parameters of the field as the energy of the microcanonical field simulation is varied.
\end{abstract}


\date{\today}

\maketitle

%%%%%%%%%%%%%%%%%%%%%%%%%%%%%%%%%%%%%%%%%%%%%%%%%%%%%%%%%%%%%%%%%%%%%%%%%%%%%%%%%%%%%%%%%%%%%%%%%%%%%%%%%%%%%%%%%%%%%%%%%%%%%%%%%%%%
\section{Introduction}\label{sec:Introduction}
Since the first experimental observations \cite{Anderson95,Davis95,Bradley95} of Bose-Einstein condensation in dilute alkali gases, much interest has been directed to the study of quantum vortices in such systems, both experimentally \cite{Madison00,Anderson00,Abo-Shaeer01} and theoretically (for reviews see \cite{Fetter01,Parker08,Kasamatsu08,Fetter09}).  On one hand, dilute condensates offer an opportunity to gain insight into the physics of superfluids and vortices familiar from condensed-matter systems.  The weakly interacting nature of dilute condensates makes them amenable to \emph{tractable} theoretical approaches, complementing the precision of control and observation afforded by dilute atomic-gas experiments \cite{Anderson00}.  However, there are also marked differences between the vortex physics of dilute condensates and that of bulk superfluids.  In contrast to bulk superfluids, dilute condensates have inhomogeneous density profiles, leading to new vortex physics \cite{Fetter01,Sheehy04,Mason08}.  Furthermore the behavior of dilute gas condensates is in many cases subtly different to that expected from experience with bulk superfluids: examples are the much slower collision rates in the dilute condensates, resulting in the breakdown of two-fluid hydrodynamics \cite{Leggett01, Zaremba99}, and the essential role of dynamical instabilities in the formation of vortex lattices in stirred condensates \cite{Sinha01}.  

Early numerical investigations by Butts and Rokhsar \cite{Butts99} revealed some of the the nature of the rotating condensates at zero temperature.  As in bulk superfluids \cite{Hess67}, dilute condensates seek to mimic rigid-body rotation \cite{Feder01} by acquiring vortices in a regular array.  Condensates adjust their rotation both by undergoing discontinuous phase transitions \cite{Butts99,Vorov05} between rotational-symmetry classes by nucleating additional vortices, and also by adjusting their vortex distributions within a symmetry class.  At a critical imposed rotation frequency, the ground state of the system makes a transition from an irrotational, vortex-free state to a central-vortex state \cite{Dalfovo99}, with angular momentum per atom $\ell\equiv\langle L_z\rangle/N\hbar=1$.  This transition is continuous, and the system possesses a continuum of off-axis vortex states with $0<\ell<1$.  Unlike the stationary $\ell=0$ and $\ell=1$ states, these intermediate states are not ground states in any rotating frame \cite{Butts99}, and are thermodynamically unstable in all frames.  However, they are dynamically stable, and at zero temperature the vortex follows a stable orbit about the trap center.  Such states represent the simplest nontrivial example of vortex dynamics in BEC.  The precession of such a vortex can be viewed as the vortex being carried by its own \emph{self-induced} superflow \cite{Donnelly91}, arising due to the effective boundary conditions imposed on the superflow by the inhomogeneity of the condensate orbital density \cite{Guilleumas01}.  Corrections to this picture due to the finite extent of the vortex core have been investigated in \cite{Nilsen06} and \cite{Jezek08}.

Further complexity is attained in finite-temperature scenarios, where the vortex may be subject to additional forces resulting from its interaction with the thermal field \cite{Sonin97, Thouless96}.  If the angular velocity of the thermal cloud is different from that of the vortex, the vortex experiences a frictional force which causes it to drift radially.  For thermal-cloud angular velocities below some threshold \cite{Fetter01}, this interaction induces the expulsion of the vortex from the condensate \cite{Fedichev99}.  The dynamics of a vortex at finite temperature are of interest both as a test of theories of finite-temperature condensates \cite{Schmidt03,Isoshima04,Jackson09,Blakie08}, and for understanding the nature of vortex interactions with bulk thermal flows, which has proved a controversial issue \cite{Sonin97,Thouless96,Flaig06}. In this paper we realize finite-temperature precessing single-vortex states as \emph{equilibrium} configurations of a conserving (Hamiltonian) classical-field system \cite{Blakie08}, with fixed internal energy and angular momentum.  Such Hamiltonian classical-field methods and closely related stochastic-field methods \cite{Gardiner03,Bradley08} have proven useful both in equilibrium and nonequilibrium scenarios \cite{Blakie05,Davis06,Simula06,Wright08,Bezett09,Weiler08,Blakie08}.  Working at fixed angular momentum allows us to create rotational-symmetry-broken precessing-vortex states at finite-temperature equilibrium of the classical field.  In such configurations, the vortex is at rotational equilibrium with the thermal cloud, and the dissipative force responsible for vortex decay \cite{Fedichev99,Madarassy08,Jackson09} vanishes.  We extract the symmetry-broken condensate orbital from the classical-field equilibrium, and observe both the Goldstone mode associated with breaking rotational symmetry, and the filling of the vortex core by the thermal component of the field.  Increasing the field energy at fixed angular momentum we observe an increase in vortex precession radius and frequency, ultimately leading to the expulsion of the vortex from the condensate and the decoupling of the condensate rotation from that of the thermal cloud. 

This paper is organized as follows:  In Sec.~\ref{sec:Formalism} we introduce the formalism we use in this work, and discuss its interpretation.  In Sec.~\ref{sec:Simulation_procedure}, we discuss our simulation procedure, introducing in turn the parameters of our system, and the details of our numerical approach.  In Sec.~\ref{sec:Results} we present results of representative simulations, and discuss techniques used in the extraction of physical information from the simulation trajectories.  In Sec.~\ref{sec:Energy_dependence} we discuss the dependence of the system behavior on the internal energy of the field, and in Sec.~\ref{sec:Conclusions} we summarize and present our conclusions.  
%%%%%%%%%%%%%%%%%%%%%%%%%%%%%%%%%%%%%%%%%%%%%%%%%%%%%%%%%%%%%%%%%%%%%%%%%%%%%%%%%%%%%%%%%%%%%%%%%%%%%%%%%%%%%%%%%%%%%%%%%%%%%%%%%%%%
\section{Formalism}\label{sec:Formalism}
%%%%%%%%%%%%%%%%%%%%%%%%%%%%%%%%%%%%%%%%%%%%%%%%%%%%%%%%%%%%%%%%%%%%%%%%%%%%%%%%%%%%%%%%%%%%%%%%%%%
\subsection{Review of Hamiltonian classical field method}
In this paper we employ classical-field-theory techniques, which have recently been reviewed at length in \cite{Blakie08}.  Here we recapitulate the basic elements of the Hamiltonian classical field method, and briefly discuss the appropriate interpretation of its application here.

At its simplest level, classical-field theory results from the replacement of a second-quantized (Bose) field Hamiltonian with its classical analogue.  This is motivated by the observation that in regimes of thermal behavior, the dominant characteristic of the atomic Bose-field system is its multimode, self-interacting nature.  Indeed,  under conditions of high mode occupation, field commutators become relatively unimportant, and a satisfactory description of the bosonic field can be obtained from a classical-field model.  More generally, the dynamics of the classical-wave system itself are both of intrinsic interest \cite{Connaughton05}, and able also to provide qualitative insight into the \emph{dynamics} of degenerate Bose-gas systems in scenarios where a precise identification of the method with the many-body field theory is impractical \cite{Wright08}.  In the present work we consider a closed, Hamiltonian system, with fixed angular momentum.  This allows us to characterize the motion of a quantum vortex \emph{at equilibrium} with a rotating thermal component of the field.

The essential ingredient of the classical-field method as developed and refined by Davis, Blakie and co-workers \cite{Davis02,Blakie05,Gardiner03,Blakie08} is the projection operator.  While the classical dynamics of infinite-dimensional systems is able to be perfectly well-defined (see e.g. \cite{Sudarshan74}), the projection operator restricts the system to a finite number of degrees of freedom \cite{noteA}.  This projection reduces the dynamics of the field to a finite Hamiltonian system, suitable for numerical implementation. Furthermore it acts to regularize the divergences resulting from the contact scattering potential assumed in the field theory \cite{Bruun99,Morgan00}, such that a formal renormalization can then be performed if so desired \cite{Norrie06a}. 

The finite-dimensional Hamiltonian system so obtained exhibits ergodic behavior \cite{Lebowitz73}, and this provides a useful tool for analyzing the equilibrium properties of Bose-gas systems.  By time-averaging functionals of the projected field along a sufficiently long field trajectory, one expects to obtain their mean values in the microcanonical ensemble of field configurations satisfying the constraints of the conserved first integrals of the system \cite{Sethna06}.  These mean values then serve as estimates of the analogous quantum-statistical-mechanical correlations of the many-body quantum system.  With a careful choice of projector, the classical (equipartitioned) equilibrium distribution can be chosen to be coincident with the high-occupation limit of the true bosonic system, allowing for highly accurate estimates to be made for the quantum system \cite{Davis06}. 

The utility of the classical-field method in its application to dynamical systems is its ability to include the effects of such \emph{thermal} behavior \emph{nonperturbatively} in the description of the `macroscopic' dynamics of the field \cite{Wright08,Bezett09}.  The coherent fraction's evolution is influenced by the (locally) thermal behavior of the incoherent fraction, including both mean-field repulsion and dissipative effects in its resulting dynamics.  The coupled dynamics of the two components are described also in the method of \cite{Zaremba99}, with the crucial difference that in the classical-field approach no distinction between condensed and thermal material is made in deriving equations of motion or implementing their solution numerically.  Instead the condensate is identified \emph{a posteriori} from the fluctuation statistics of the field.  

In this work, while we wish to study the behavior of the classical-field system at equilibrium, there is an essential \emph{dynamical} aspect to the equilibrium, i.e., the precession of the vortex.  Our interest is in the dynamics of the vortex, and thus of the condensate, in the presence of the thermal component of the field.  We thus follow \cite{Bezett09} in employing the classical-field method to include the effects of the thermal component of the field on the motion of a collective excitation of the condensate.  In contrast to that work the collective excitation we study here (the precession of the vortex) is stable at thermal equilibrium of the system.  The equilibrium trajectory of the vortex, which is determined by the balance of coherent and thermal forces acting on it, defines the frame in which the rotational-symmetry-broken condensate mode is stationary.  The ergodic character of the field evolution tends to restore this broken rotational symmetry as time progresses, and we are thus lead to abandon formal ergodic averages and consider the coherence properties of the field on short time scales, as we discuss in Sec.~\ref{sec:Results}.
%%%%%%%%%%%%%%%%%%%%%%%%%%%%%%%%%%%%%%%%%%%%%%%%%%%%%%%%%%%%%%%%%%%%%%%%%%%%%%%%%%%%%%%%%%%%%%%%%%%
\subsection{System}\label{subsec:System}
We consider here a system harmonically trapped with a trapping potential isotropic in the $xy$~plane:
\begin{equation}
	V(\mathbf{x}) = \frac{m}{2}\Big[\omega_r^2(x^2+y^2) + \omega_z^2z^2\Big].
\end{equation}
We choose highly oblate harmonic trapping, with trapping frequency $\omega_z$ sufficiently high that no modes are excited in the $z$-direction.  A two-dimensional (2D) description is thus valid provided that \cite{Petrov00},
\begin{equation}
	\mu+k_\mathrm{B}T \ll \hbar\omega_z,
\end{equation}
where $\mu$ and $T$ are the system chemical potential and temperature respectively (see also the discussion in Ref.~\cite{Wright08}).  Restricting the system to 2D permits the dynamics of a \emph{point} vortex interacting with the thermal component of the field to be studied, while removing complexities such as line bending and Kelvin-wave excitations of a vortex filament of finite extent, which provide additional mechanisms for vortex dissipation \cite{Kozik04}.  Following Ref.~\cite{Wright08}, we choose physical parameters corresponding to $^{23}\mathrm{Na}$ atoms confined in a strongly oblate trap, with trapping frequencies $(\omega_r,\omega_z)=2\pi\times(10,2000)$ rad/s.  The s-wave scattering length is $a=2.75$nm, placing our system in the quasi-2D regime ($l_z \equiv \sqrt{\hbar/m\omega_z} \gg a$), with an effective 2D interaction parameter $U_\mathrm{2D}=2\sqrt{2\pi}\hbar a/ml_z$.  In our simulations we will often compare our systems to what we will refer to as our principal \emph{ground state}, which we take as the ground Gross-Pitaevskii (GP) eigenmode \cite{Dalfovo99} of the trapped system in a frame rotating at angular velocity $\Omega_0=0.35\omega_r$, with eigenvalue $(\mu_\mathrm{g})_{\Omega_0}=10\hbar\omega_r$ in that frame. This state consists of $N_0=1.072\times10^4$ atoms, and contains a single on-axis vortex carrying unit angular momentum per atom (angular momentum and rotation will always refer to that about the $z$~axis in this paper).  In an inertial (non-rotating) frame this state has eigenvalue $\mu_\mathrm{g}=10.35\hbar\omega_r$, and energy $E_\mathrm{g}=7.646\times10^4\hbar\omega_r$.  In this paper we will refer generically to any such inertial frame as a laboratory frame (lab frame).
%%%%%%%%%%%%%%%%%%%%%%%%%%%%%%%%%%%%%%%%%%%%%%%%%%%%%%%%%%%%%%%%%%%%%%%%%%%%%%%%%%%%%%%%%%%%%%%%%%%
\subsection{Equations of motion}
To derive the classical-field equation of motion, we begin with the second-quantized Hamiltonian of a trapped Bose gas in the $s$-wave regime
\begin{equation}\label{eq:FullH}
H=\int d^3\mathbf{x}\; \hat{\Psi}^\dag(\mathbf{x}) H_{\mathrm{sp}} \hat{\Psi}(\mathbf{x}) +\frac{U}{2} \int d^3\mathbf{x}\;\hat{\Psi}^\dag(\mathbf{x})  \hat{\Psi}^\dag(\mathbf{x}) \hat{\Psi}(\mathbf{x})\hat{\Psi}(\mathbf{x}), 
\end{equation}
where $\hat{\Psi}(\mathbf{x})$ is the bosonic field operator satisfying $[\hat{\Psi}(\mathbf{x}),\hat{\Psi}^\dagger(\mathbf{x}')]=\delta(\mathbf{x}-\mathbf{x}')$, and the interaction is assumed to be a contact potential with $U=4\pi\hbar^2a/m$, $a$ the scattering length and $m$ the atomic mass.  We choose to write the Hamiltonian in a frame rotating at angular velocity $\Omega$ about the $z$~axis.  As the interaction potential is rotationally invariant this choice affects only the definition of the single-particle Hamiltonian
\begin{equation}\label{eq:sp_Hamiltonian}
	H_\mathrm{sp} = \frac{-\hbar^2\nabla^2}{2m} + \frac{m}{2}\Big(\omega_r^2r^2+\omega_z^2z^2\Big) - \Omega L_z.
\end{equation}
We introduce a cutoff energy $E_R$, delimiting a low-energy region ($\mathbf{L}$) of the system consisting of single-particle modes [eigenmodes of Eq.~(\ref{eq:sp_Hamiltonian})] with single-particle energies  $\epsilon_n\leq E_R$, which is sometimes referred to as the \emph{condensate band} \cite{Blakie08}.  The remaining modes comprise the complementary high-energy region (noncondensate band $\mathbf{H}$).  The cutoff $E_R$ is chosen to be at a sufficiently high energy that the interacting Hamiltonian Eq.~(\ref{eq:FullH}) is approximately diagonal at the cutoff \cite{Gardiner03}.  This division is enforced formally by the projection operator defined
\begin{equation}\label{eq:Pdef} 
{\cal P}f(\mathbf{x})\equiv\sum_{n \in
\mathbf{L}}\phi_n(\mathbf{x})\int d^3\mathbf{y}\;
\phi_n^*(\mathbf{y})f(\mathbf{y}),  
\end{equation} 
where summation is over all single-particle modes satisfying $\epsilon_n \leq E_R$.  We refer to the number of such modes spanning the low-energy region as the condensate-band multiplicity ($\mathcal{M}$).  We use this projector to define the projected field operator
\begin{equation}\label{eq:proj_field_opr}
\hat{\psi}(\mathbf{x}) \equiv  \sum_{n \in\mathbf{L}}\hat{a}_n\phi_n(\mathbf{x})= {\cal P}\hat{\Psi}(\mathbf{x}).   
\end{equation}
We choose the cutoff such that $E_R<\hbar\omega_z$, and thus all excited $z$-axis modes are excluded from the low-energy basis over which $\hat{\psi(\mathbf{x})}$ is expanded.  A classical theory is obtained by demoting the operators $\hat{a}_n$ in Eq.~(\ref{eq:proj_field_opr}) to classical variables $\alpha_n$, thereby defining a projected classical field $\psi(\mathbf{x}) = \sum_{n\in\mathbf{L}}\alpha_n\phi_n(\mathbf{x})$.  The corresponding classical-field Hamiltonian is
\begin{equation}\label{eq:HCF}
H_{\mathrm{CF}}=\int d^2\mathbf{x}\; \psi^*(\mathbf{x}) H_{\mathrm{sp}}
\psi(\mathbf{x}) +\frac{U_\mathrm{2D}}{2} |\psi(\mathbf{x})|^4,  
\end{equation}
which is formally a classical Hamiltonian with $2\mathcal{M}$ canonical degrees of freedom $\{P_n,Q_n\}$ related to the field variables $\{\alpha_n,\alpha_n^*\}$ by a canonical transformation \cite{Davis03}.  It then follows that the Hamilton's equations can be expressed in terms of projected functional differentiation of the classical-field Hamiltonian \cite{Gardiner03} as
\begin{equation}\label{eq:PGPE} 
i\hbar\frac{\partial\psi(\mathbf{x})}{\partial
t}=\frac{\bar{\delta}H_{\mathrm{CF}}}{\bar{\delta}\psi^*(\mathbf{x})}
= {\cal P}\left\{\left(H_{\mathrm{sp}}+U_\mathrm{2D}|
\psi(\mathbf{x})|^2\right)\psi(\mathbf{x})\right\}, 
\end{equation} 
which is the projected Gross-Pitaevskii equation (PGPE) (see, e.g., \cite{Blakie08}).  The conservation of field energy $E[\psi]=H_\mathrm{CF}$ and normalization $N=\int d^2\mathbf{x}|\psi(\mathbf{x})|^2$ under the action of the PGPE follow immediately from the nature of $H_\mathrm{CF}$ as a time-invariant classical Hamiltonian, and its invariance under the transformation $\psi(\mathbf{x})\rightarrow \psi(\mathbf{x})e^{i\theta}$ respectively \cite{Sudarshan74}.  Furthermore the rotational invariance of the Hamiltonian $H_\mathrm{CF}$ implies that the angular momentum 
\begin{equation}\label{eq:ang_mom}
	\overline{L_z} = \int d^2\mathbf{x} \psi^*(\mathbf{x})L_z\psi(\mathbf{x}) 
\end{equation}
[where the bar denotes spatial averaging: $\overline{A}=\int d^2 \mathbf{x}\psi^*(\mathbf{x})A\psi(\mathbf{x})$] is also conserved by the evolution of the PGPE. 
%%%%%%%%%%%%%%%%%%%%%%%%%%%%%%%%%%%%%%%%%%%%%%%%%%%%%%%%%%%%%%%%%%%%%%%%%%%%%%%%%%%%%%%%%%%%%%%%%%%%%%%%%%%%%%%%%%%%%%%%%%%%%%%%%%%%
\section{Simulation procedure}\label{sec:Simulation_procedure}
%%%%%%%%%%%%%%%%%%%%%%%%%%%%%%%%%%%%%%%%%%%%%%%%%%%%%%%%%%%%%%%%%%%%%%%%%%%%%%%%%%%%%%%%%%%%%%%%%%%
\subsection{Numerical implementation}\label{subsec:Numerical implementation}
\subsubsection{Dimensionless units}
For the purpose of the implementation of the PGPE, it is convenient to express the physical quantities $\{r,\omega,E,t\}$ in a dimensionless form, which we write as $\{\bar{r},\bar{\omega},\bar{E},\bar{t}\}$.  These quantities are related by the expressions $r=\bar{r}r_0$, $\omega=\bar{\omega}\omega_r$, $E=\bar{E}\hbar\omega_r$, $t=\bar{t}\omega_r^{-1}$, where the radial oscillator length $r_0=\sqrt{\hbar/m\omega_r}$.  In reporting our results we will often refer to times in units of the trap cycle (abbreviated cyc.), i.e., the period of the radial oscillator potential $T_\mathrm{osc}=2\pi/\omega_r$.   
\subsubsection{PGPE implementation}\label{subsubsec:PGPE_implementation}
Here we briefly discuss our implementation of the PGPE in terms of rotating-frame harmonic-oscillator states, as described in \cite{Bradley08,Wright08}.  In dimensionless form the PGPE [Eq.~(\ref{eq:PGPE})] becomes
\begin{equation}\label{eq:pgpeBar}
	i\frac{\partial\bar{\psi}}{\partial \bar{t}} =
	\mathcal{P}\Bigg\{\Big[-\frac{\bar{\nabla}^2}{2} +
	\frac{\bar{r}^2}{2} + 
	i\bar{\Omega}\partial_\theta + \lambda|\bar{\psi}|^2\Big]\bar{\psi}\Bigg\},
\end{equation} 
where the dimensionless effective interaction strength $\lambda = U_\mathrm{2D}/r_0^2\hbar\omega_r$.
We proceed as in \cite{Bradley08} by
expanding $\bar{\psi}$ as \begin{equation}\label{eq:expansion}
	\bar{\psi}(\bar{\mathbf{x}},\bar{t}) = \sum_{\{n,l\}}
	c_{nl}(\bar{t})\bar{Y}_{nl}(\bar{r},\theta),
\end{equation} over the Laguerre-Gaussian modes
\begin{equation}\label{eq:Laguerre-Gaussian}
	\bar{Y}_{nl}(\bar{r},\theta) = \sqrt{\frac{n!}{\pi(n+|l|)!}}
	e^{il\theta}\bar{r}^{|l|}e^{-\bar{r}^2/2}L_n^{|l|}(\bar{r}^2),
\end{equation} which diagonalize the single-particle (non-interacting)
Hamiltonian $\bar{H}_\mathrm{sp}^0$ \begin{equation}
	\bar{H}_\mathrm{sp}^0\bar{Y}_{nl}(\bar{r},\theta) =
	\Big[-\frac{\bar{\nabla}^2}{2} + \frac{\bar{r}^2}{2} +
	i\bar{\Omega}\partial_\theta\Big]\bar{Y}_{nl}(\bar{r},\theta) =
	\bar{E}_{nl}^\Omega \bar{Y}_{nl}(\bar{r},\theta),
\end{equation} with eigenvalues $\bar{E}_{nl}^\Omega = 2n + |l| - \bar{\Omega} l
+ 1$.  The energy cutoff defined by the projector $\mathcal{P}$ requires the
expansion of Eq.~(\ref{eq:expansion}) to exclude all terms except those in
oscillator modes $\bar{Y}_{nl}(\mathrm{r},\theta)$ for which
$\bar{E}_{nl}^\Omega \leq \bar{E}_R$, i.e.  those modes satisfying
\begin{equation}\label{eq:band_spectrum}
	2n+|l| - \bar{\Omega} l + 1 \leq \bar{E}_R.
\end{equation} 
This yields the equation of motion
\begin{equation}\label{eq:PGPE_coeff_eom}
	i\frac{\partial c_{nl}}{\partial \bar{t}} = \bar{E}_{nl}^\Omega c_{nl} +
	\lambda F_{nl}(\bar{\psi}).
\end{equation} 
where the projection of the GP nonlinearity
\begin{equation}\label{eq:nonlinearity_quadrature}
	F_{nl}(\bar{\psi}) = \int_0^{2\pi} d\theta \int_0^\infty \bar{r}d\bar{r}
	\bar{Y}^*_{nl}(\bar{r},\theta)|
	\bar{\psi}(\bar{r},\theta)|^2\bar{\psi}(\bar{r},\theta),
\end{equation} 
is evaluated by the technique presented in \cite{Bradley08}.
%%%%%%%%%%%%%%%%%%%%%%%%%%%%%%%%%%%%%%%%%%%%%%%%%%%%%%%%%%%%%%%%%%%%%%%%%%%%%%%%%%%%%%%%%%%%%%%%%%%
\subsection{Microcanonical evolution}\label{subsec:Microcanonical_evolution}
%%%%%%%%%%%%%%%%%%%%%%%%%%%%%%%%%%%%%%%%%%%%%%%%%%%%%%%%%%%
\subsubsection{Microcanonical formalism}\label{subsubsec:Microcanonical_formalism}
Davis, Blakie and co-workers have shown \cite{Davis01,Blakie05} that the PGPE can be used to evolve a general configuration of the classical field to a thermal equilibrium.  In contrast to those works, in which the initial field configurations were parameterized solely by their energies
\begin{equation}\label{eq:conserved_energy}
	E[\psi]\equiv H_{\mathrm{CF}}[\psi] =\int d^2\mathbf{x}\; \psi^*(\mathbf{x}) H_{\mathrm{sp}}  \psi(\mathbf{x}) +\frac{U_\mathrm{2D}}{2} |\psi(\mathbf{x})|^4, 
\end{equation}
in this work we also fix the conserved angular momentum 
\begin{equation}\label{eq:conserved_ang_mom}
	L[\psi] \equiv \int d^2\mathbf{x}\; \psi^*(\mathbf{x})\hat{L_z}\psi(\mathbf{x}),
\end{equation}
to a prescribed nonzero value \cite{noteB}.  We therefore expect that the field evolves under the action of the PGPE to a thermal equilibrium consistent with the choices of conserved energy [Eq.~(\ref{eq:conserved_energy})] and angular-momentum [Eq.~(\ref{eq:conserved_ang_mom})] first integrals.  In this paper we choose $L[\psi]=N[\psi]\hbar=N_0\hbar\equiv L_0$, and thus explore the manifold of classical-field configurations with unit angular momentum per particle.  At zero temperature (corresponding to complete condensation in the classical-field model employed here, in which no representation of quantum depletion is included), the corresponding microstate (unique up to a choice of phase) is our principal ground state, containing a singly charged, on-axis vortex, with energy $E[\psi]=E_\mathrm{g}$.  By choosing initial configurations with $E[\psi]>E_\mathrm{g}$ and $L_z[\psi]=L_0$, we investigate the thermal equilibria that result when atoms are thermally excited out of the ground condensate mode, while conserving the total angular momentum. 
%%%%%%%%%%%%%%%%%%%%%%%%%%%%%%%%%%%%%%%%%%%%%%%%%%%%%%%%%%%
\subsubsection{Choice of frame}\label{subsubsec:Choice_of_frame}
In applying our classical-field method, we must make a choice of the rotation rate of the frame in which the PGPE is defined, which we will denote by $\Omega_\mathrm{p}$ in the remainder of this paper.  The only non-trivial consequence of this choice \cite{noteV} is the precise definition of the projector $\mathcal{P}=\mathcal{P}(E_R;\Omega_\mathrm{p})$, and thus of the set of modes which span the low-energy space $\mathbf{L}$.  This dependence of the projector on $\Omega_\mathrm{p}$ generalizes the familiar concept of an energy cutoff in classical-field theory.  As is well known \cite{Blakie08}, the thermodynamic parameters of the classical field at equilibrium depend both on the conserved first integrals of the system and the imposed cutoff.  In the present work, we expect that the equilibrium of the system minimizes the free energy 
\begin{equation}\label{eq:free_energy}
	F = E - TS - \Omega L,
\end{equation}
where the thermodynamic angular velocity $\Omega=(\partial E/\partial L)_S$ \cite{Landau69}.  The utility of the rotationally invariant (Gauss-Laguerre) projector used here is that the classical field evolution is not \emph{dynamically} biased towards a particular rotating frame \cite{Wright08}, i.e., we retain a (generalized) Ehrenfest relation \cite{Bradley05} 
\begin{equation}\label{eq:Ehrenfest_Lz}
	\frac{d\overline{L_z}}{dt} = -\frac{i}{\hbar}\overline{L_zV(\mathbf{x})},
\end{equation}
for the angular momentum $\overline{L_z}$.  This ensures that for the isotropic trapping we consider here, angular momentum is conserved, and further that the field is free to rotate at the angular velocity $\Omega$ which minimizes Eq.~(\ref{eq:free_energy}), which will in general \emph{not} be equal to the projector angular velocity $\Omega_\mathrm{p}$.  The relationship between the first integrals $(E,L)$ and the thermodynamic variables $(T,\Omega)$ of course depends on $\Omega_\mathrm{p}$, as different choices of cutoff yield different Hamiltonian systems.  We demonstrate this dependence by explicit calculation in Appendix~\ref{app:AppendixA}, and note that the use of a harmonic basis allows the PGPE theory to be extended to include a mean-field description of above-cutoff atoms, whereby insensitivity to (moderate) changes in the cutoff is regained (see \cite{Blakie08}).  However, as a first application to the precessing-vortex system, we restrict our attention to a strictly Hamiltonian ``PGPE system'' \cite{Blakie05}, with low-energy region $\mathbf{L}$ defined by application of the projector in the frame in which the ground state was formed (i.e., we take $\Omega_\mathrm{p}=\Omega_0=0.35\omega_r$), and investigate the behavior of this particular Hamiltonian system by varying its energy at fixed angular momentum.
%%%%%%%%%%%%%%%%%%%%%%%%%%%%%%%%%%%%%%%%%%%%%%%%%%%%%%%%%%%
\subsubsection{Procedure}\label{subsubsec:Procedure}
For almost any initial configuration far from equilibrium, we expect the field to approach equilibrium over time \cite{Lebowitz73}, due to the ergodicity of the system.  
We thus form our initial states as randomized field configurations under the projector defined by cutoff energy $E_R=3(\mu_\mathrm{g})_{\Omega_0}=30\hbar\omega_r$ and rotation frequency $\Omega_\mathrm{p}=0.35\omega_r$, with the same values of normalization $N$ and angular momentum $L$ as our principal ground state, but with higher energies.  These states are essentially formed by `mixing' a randomized high-energy state with the principal ground state in a procedure similar to that described in \cite{Blakie08}.
The initial-state information will persist in the precise details of the fluctuations \cite{Connaughton05} since the field equations are deterministic, although reversibility of the evolution is lost after some time to numerical error \cite{Lebowitz73}.  
The values of the conserved first integrals $L$ and $E$ are fixed to the desired values to within specified tolerances ($\Delta L/L_0 = 10^{-6}, \Delta E/E_\mathrm{g} = 10^{-4}$).  We form initial states in this manner with a range of energies.  In each case the central vortex of the principal ground state remains, however the density profile of each state is severely distorted.  As all our states have $L = N_0\hbar$, the inertial-frame and projector-frame energies are in each case related by a constant shift resulting from the fixed rotational kinetic energy of the field: $(E)_0 = (E)_{\Omega_\mathrm{p}} + \hbar\Omega_\mathrm{p} N_0$.  This shift is of order $5\%$ for the ground-state wavefunction.  In this paper we will discuss the energies of states in the \emph{lab frame}, and quote them as a multiple of the ground-state energy $E_\mathrm{g}$ in that frame. 

Having formed the initial states we evolve them in the projector frame for a period of $10^4$ trap cycles, with an adaptive integrator \cite{Davis_DPhil} with accuracy chosen such that the relative change in field normalization is $\leq 4\times10^{-9}$ per time step taken \cite{Wright08}.  By $9\times10^3$ trap cycles the states have reached equilibrium, as evidenced by the settling of the distribution of particles in momentum space \cite{Davis01,noteD}, and we perform our analysis of steady-state properties on the final $10^3$ trap cycles of evolution data. 
%%%%%%%%%%%%%%%%%%%%%%%%%%%%%%%%%%%%%%%%%%%%%%%%%%%%%%%%%%%%%%%%%%%%%%%%%%%%%%%%%%%%%%%%%%%%%%%%%%%
%%%%%%%%%%%%%%%%%%%%%%%%%%%%%%%%%%%%%%%%%%%%%%%%%%%%%%%%%%%%%%%%%%%%%%%%%%%%%%%%%%%%%%%%%%%%%%%%%%%%%%%%%%%%%%%%%%%%%%%%%%%%%%%%%%%%
\section{Results}\label{sec:Results}
\begin{figure*}
	\includegraphics[width=0.9\textwidth]{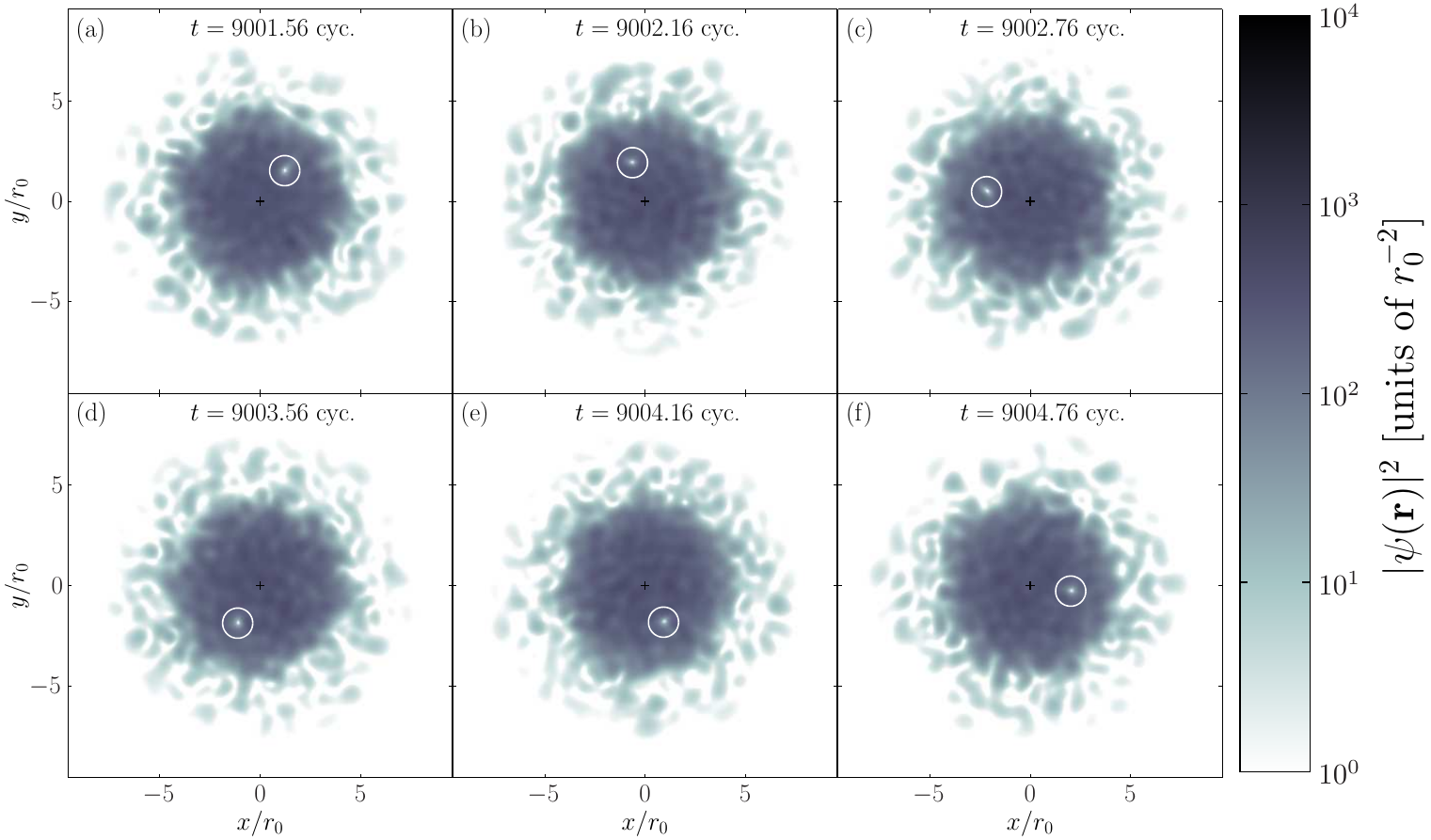}
	\caption{\label{fig:density_plots1} (Color online) Classical-field density during a single orbit of the vortex, as viewed in the lab frame.  The white circle indicates the vortex position, and $+$ marks the coordinate origin (trap axis).  Parameters of the classical field are given in the text.}
\end{figure*}

After the equilibration period, the behavior of the classical field simulations is qualitatively similar for energies in the range $E=[1.04,1.15]E_\mathrm{g}$.  Each exhibits a central high-density region containing a vortex which is displaced off and precessing about the trap axis.  Viewed in the lab frame, the precession is in the same sense as the vortex (i.e. positive rotation sense, corresponding to angular-momentum numbers $m > 0$), and is stochastic in nature, with the vortex position fluctuating about a circular mean orbit, in the presence of strong density fluctuations of the background field.  As noted in Sec.~\ref{subsubsec:Choice_of_frame}, the frequency of the vortex precession is in general not equal to the angular velocity of the projector defining the microcanonical system. The condensate is located entirely in the central bulk (see Sec.~\ref{subsec:Penrose_Onsager_analysis}); outside this region the field exhibits the high-energy turbulence \cite{Wright08} characteristic of purely thermal atoms in the classical-field model.  In Fig.~\ref{fig:density_plots1}(a-f) we plot classical-field densities spanning a single orbit of the vortex in a simulation with conserved energy $E=1.10E_\mathrm{g}$.  The behavior presented in Fig.~\ref{fig:density_plots1} is representative of all simulations in which the equilibrium condensate contains a precessing vortex.
From a range of simulations we observe that the radius at which the vortex precesses increases as the energy of the field is increased.  At higher energies it approaches the edge of the condensate and at $E\approx1.16E_\mathrm{g}$ it is ultimately lost into the violently evolving peripheral region of surface excitations and short-lived phase defects.  As the field energy is decreased, the vortex approaches the trap center.  However, we expect that as the energy of the field is lowered towards the ground state, the excitation becomes too weak to reliably generate thermal statistics on the time scales of interest, and we suspect that for low energies the background field `seen' by the vortex may not appear thermal on the time scale of its precession.  Indeed in such a low temperature regime the effects of quantum fluctuations neglected in our model will become important.  We therefore exclude simulations with energies $E<1.04E_\mathrm{g}$ from our analysis.  A detailed discussion of the dependence of system observables on the internal energy of the field is presented in Sec.~\ref{sec:Energy_dependence}.
%%%%%%%%%%%%%%%%%%%%%%%%%%%%%%%%%%%%%%%%%%%%%%%%%%%%%%%%%%%%%%%%%%%%%%%%%%%%%%%%%%%%%%%%%%%%%%%%%%%
\subsection{Vortex precession}\label{subsec:Vortex_precession}
\begin{figure}
	\includegraphics[width=0.45\textwidth]{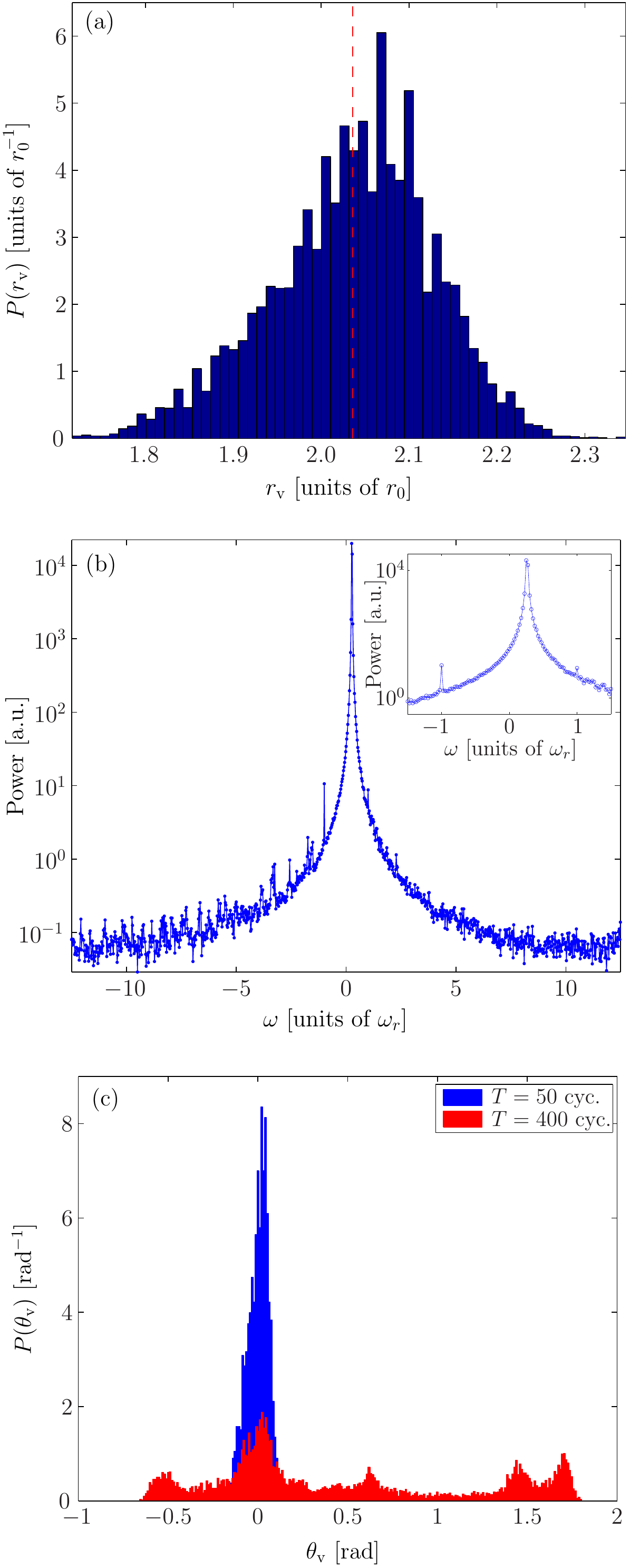}
	\caption{\label{fig:equil_vortex_motion} (Color online) Quantities characterizing the motion of the vortex. (a) Normalized histogram of measured vortex displacement magnitudes.  The vertical dashed line indicates the mean. (b) Power spectrum of the complex vortex coordinate $z_\mathrm{v} \equiv x_\mathrm{v} + iy_\mathrm{v}$, revealing the precession of the vortex as viewed in the laboratory frame. (c) Normalized histograms of vortex phases $\theta_\mathrm{v}$ measured in a rotating frame (see main text) over periods of length 50 cyc.\ and 400 cyc., starting from t=9000 cyc.}
\end{figure}

%%%%%%%%%%%%%%%%%%%%%%%%%%%%%%%%%%%%%%%%%%%%%%%%%%%%%%%%%%%
In order to characterize the vortex motion, we track the vortex location as observed in the laboratory frame.  In practice we recorded vortex locations in the simulation with $E=1.10E_\mathrm{g}$, at a frequency of 25 samples per cycle for a period of 400 cycles, starting at $t=9000$~cyc.  We find that the displacement of the vortex from the trap center fluctuates about a mean value of $\overline{r_\mathrm{v}}=2.04r_0$, and that the series of displacements measured (from the trap center) has standard deviation $\sigma_{r_\mathrm{v}}=0.09r_0$, which is of same order as the extent of the vortex core (which is approximately given by the healing length $\eta=0.20r_0$ \cite{Dalfovo99,noteE}).  The normalized distribution of displacement radii measured over the 400 cyc.\ period is displayed in Fig.~\ref{fig:equil_vortex_motion}(a).
The plotted distribution shows some negative skew but this is not a consistent feature across the simulations performed.
To determine temporal characteristics of the vortex trajectory, we define a complex vortex coordinate $z_\mathrm{v}(t) \equiv x_\mathrm{v}(t) + iy_\mathrm{v}(t)$, where the coordinate pair $(x_\mathrm{v}(t_j),y_\mathrm{v}(t_j))$ specifies the vortex position at time $t_j$, and we calculate the power spectrum of this time series.  As direct periodograms are subject to large variances in the estimates they provide at each frequency \cite{Press92} we split the $10^4-$sample time series into 10 consecutive segments of equal length and average the power spectra obtained from these shorter series (Bartlett's method \cite{Hayes96}).  In Fig.~\ref{fig:equil_vortex_motion}(b) we plot the power spectrum of the vortex coordinate $z_\mathrm{v}$ over the (angular) frequency range $\omega \in [-12.5,12.5] \omega_r$ permitted by the sample frequency of our data \cite{noteF}.  The most conspicuous feature of the spectrum is the sharp peak [note the logarithmic scale of Fig.~\ref{fig:equil_vortex_motion}(b)] corresponding to the precession of the vortex about the trap axis.  
The next most pronounced features of the power spectrum are two clearly resolved spikes at frequencies $\omega=\pm1\omega_r$.  We identify $\omega=1\omega_r$ as the universal frequency of the dipole-oscillation mode (Kohn mode) of the harmonically trapped gas \cite{Dobson94}, which is immune to thermalization \cite{noteG}.  The frequencies $\pm1\omega_r$ thus result from a small dipole oscillation of the classical field as a whole.  Smaller features can be seen in the wings of the power-spectrum distribution, which we associate with thermal fluctuations of the classical field (c.f. discussion in \cite{Wright08}).
From the polar form $z_\mathrm{v}(t_j)=|z_\mathrm{v}(t_j)|e^{i\theta_\mathrm{v}(t_j)}$ of the vortex coordinate, we observe that the vortex phase $\theta_\mathrm{v}$ does not increase linearly with time, but fluctuates randomly as the vortex moves in the dynamically evolving thermal field.  After some time the vortex phase diffuses, so that the rotational symmetry of the system is restored in the ergodic density of classical-field configurations.  This is illustrated in Fig.~\ref{fig:equil_vortex_motion}(c), where we plot the distribution of vortex phases $\theta_\mathrm{v}$ as measured in a frame rotating at frequency $\Omega=0.2610\omega_r$.  This frequency corresponds to the peak presented in Fig.~\ref{fig:equil_vortex_motion}(b), and has been chosen because it yields the minimum variation in vortex phase over the 400 cyc.\ period we consider.  Histograms of the normalized vortex-phase distribution over the first 50 cycles of this period [blue (black) bars] and over the full 400 cyc.\ period [red (gray) bars] are plotted.  We observe that at short times the fluctuations in $\theta_\mathrm{v}$ are small and reasonably Gaussian in nature.  Over longer time periods the vortex phase drifts significantly.  This diffusion of the vortex phase produces the distinctive Lorentz-like shape of the power-spectrum distribution plotted in Fig.~\ref{fig:equil_vortex_motion}(b).  In Sec.~\ref{subsec:PO_rotating_frames} we discuss the effects of this diffusion on the `one-body' correlations of the classical field obtained from time-averaging the classical-field trajectory, and the implications for the identification and characterization of the condensate in these simulations.
%%%%%%%%%%%%%%%%%%%%%%%%%%%%%%%%%%%%%%%%%%%%%%%%%%%%%%%%%%%%%%%%%%%%%%%%%%%%%%%%%%%%%%%%%%%%%%%%%%%
\subsection{Field-covariance analysis}\label{subsec:Penrose_Onsager_analysis}
%%%%%%%%%%%%%%%%%%%%%%%%%%%%%%%%%%%%%%%%%%%%%%%%%%%%%%%%%%%%%%%%%%%%%%%%%%%%%%%%%%%%%%%%%%%%%%%%%%%
We turn now to the issue of identifying the condensate in the classical-field solutions.  In vortex-free \cite{noteH} equilibrium scenarios a method of condensate definition based on sampling of the classical-field correlations \cite{Blakie05} has proven successful.  The method, which is in obvious analogy to the Penrose-Onsager (PO) definition \cite{Penrose56,Leggett01} of Bose-Einstein condensation in terms of the (quantum) one-body density matrix, involves constructing the covariance matrix
\begin{equation}\label{eq:covar_mtx}
	\rho_{ij} = \langle c_j^*c_i \rangle_t
\end{equation}
from the propagation-basis coefficients (Sec.~\ref{subsubsec:PGPE_implementation}) of a single classical-field trajectory, where the the indices $i,j$ each index a quantum number pair $(n,l)$, and $\langle\cdots\rangle_t$ denotes an average over time. Ergodicity of the field evolution \cite{Lebowitz73} implies that averages taken over sufficiently long times approach averages over the microcanonical ensemble of field configurations, in which the coherence of a classical-field condensate \cite{Connaughton05} manifests as a single dominant eigenmode of $\rho_{ij}$.  However, this method of condensate identification has some significant ambiguities in more general scenarios, in particular when symmetries of the Hamiltonian are broken by the appropriate GP state \cite{Castin04}.  For example, the standard PO criterion itself may prove inappropriate in situations in which the condensate exhibits center-of-mass motion \cite{Wilkin98}, prompting workers to introduce revised definitions of the density matrix appropriate to to such systems, in which the center-of-mass motion is eliminated \cite{Pethick00,Yamada09}.  

Issues of condensate definition also arise in the presence of vortices.  In scenarios in which several vortices exist, breaking the rotational symmetry of an otherwise rotationally invariant many-body Hamiltonian, an uncountably infinite degeneracy of the one-body density matrix appears, even in the idealized Gross-Pitaevskii limit of a zero-temperature system \cite{Seiringer08}.  
In this paper we have chosen one of the simplest scenarios of rotational symmetry breaking in a finite-temperature classical field: the precession of a single vortex.  Because the presence of the vortex distinguishes different rotational orientations of the condensate, we find that in order to determine the condensate mode we must construct time averages in an appropriate rotating frame.  The fact that the phase of the vortex diffuses over time in any such uniformly rotating frame means that we must abandon the notion of ergodic reconstruction of the microcanonical density by time averages.  However, we expect intuitively \cite{Blakie05} that in order to quantify the condensation in the field, the averaging time need only be long enough to distinguish the Gaussian fluctuations of thermal or chaotic modes \cite{Glauber65,Mandel95} from the non-Gaussian statistics of a quasi-uniform phase evolution characteristic of a condensed component.  We find therefore that we can successfully quantify the condensation of the field from such short-time averages, while exploiting the ergodic character of the evolution over longer time scales.   In the remainder of this paper, we will use the term Penrose-Onsager (PO) procedure generically to refer to the construction of the covariance matrix (density matrix) Eq.~(\ref{eq:covar_mtx}) by a time-averaging procedure, the precise details of which will be in each case specified in the context.
%%%%%%%%%%%%%%%%%%%%%%%%%%%%%%%%%%%%%%%%%%%%%%%%%%%%%%%%%%%
\subsubsection{Lab-frame analysis}\label{subsubsec:lab-frame_PO}
\begin{figure*}
	\includegraphics[width=0.9\textwidth]{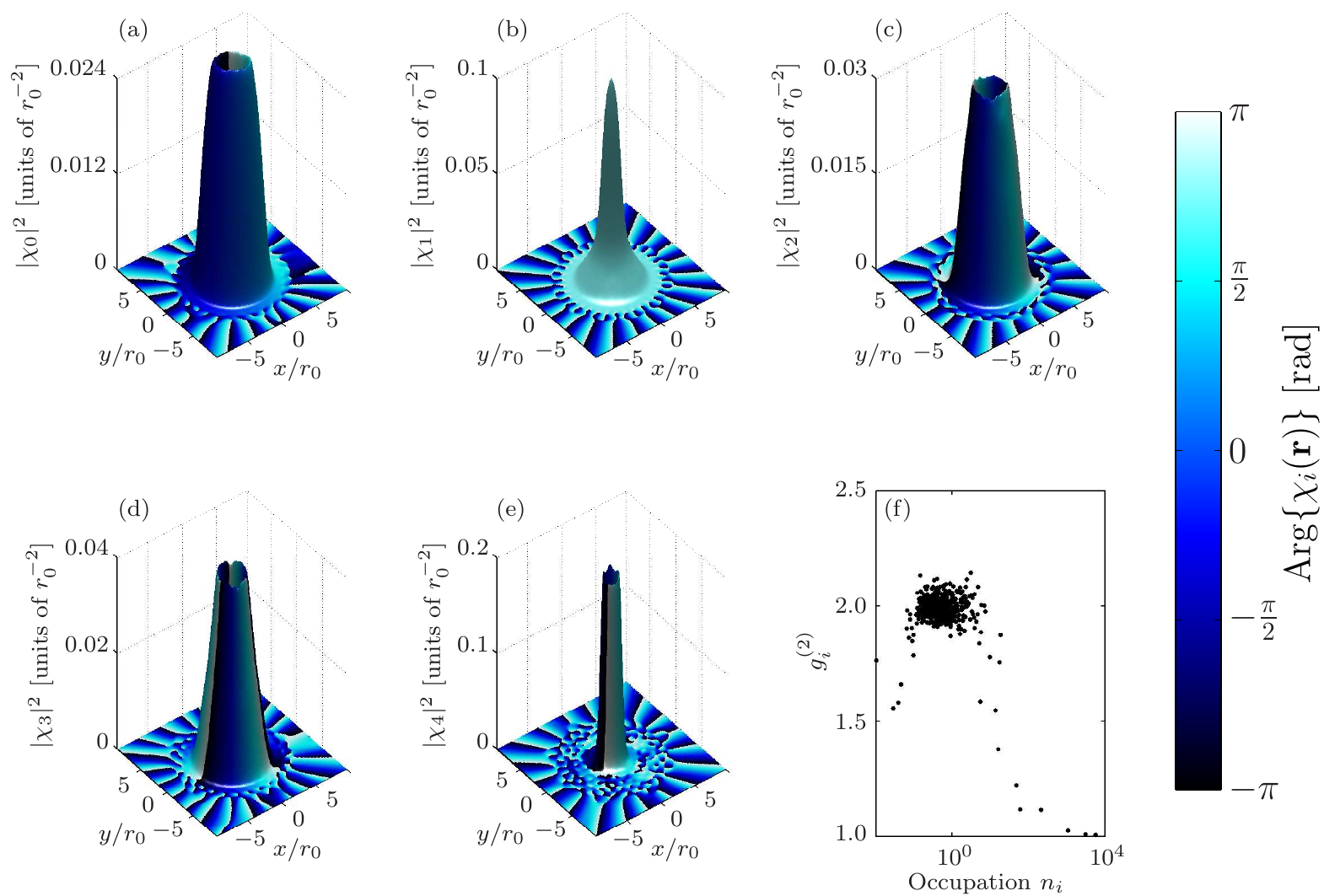}
	\caption{\label{fig:POsurfs_E105} (Color online) (a-e) Density and phase of the five most highly occupied eigenmodes of the covariance matrix, for case $E=1.05E_\mathrm{g}$. (f) Second-order coherence functions of the covariance-matrix eigenmodes versus mode occupation.}
\end{figure*}

We consider a simulation with energy $E=1.05E_\mathrm{g}$, which exhibits a vortex precessing at a frequency $\omega_\mathrm{v}\approx0.23\omega_r$, at a radius $\overline{r_\mathrm{v}}\approx1.3r_0$, which should be contrasted with the extent of the central density bulk $r_\mathrm{b}\approx5r_0$.  The vortex in this case is displaced from its axis by significantly more than the extent of its core (healing length $\eta=0.20r_0$), while the core remains comparatively close to the center of the region of significant density.  We proceed by forming the covariance matrix by averaging over $2501$ equally spaced samples of the lab-frame representation of the classical-field evolution, between $t=9000$ and $9100$~cyc.  We diagonalize the density matrix to obtain its eigenvectors, denoted by $\chi_i(\mathbf{x})$ (the eigenmodes of the one-body density operator) and their eigenvalues, denoted by $n_i$ (which are the mean \emph{occupations} of the modes during the sampled period).  The index $i$ ranges from $0$ to $\mathcal{M}-1$ in order of decreasing occupation. In Fig.~\ref{fig:POsurfs_E105}(a-e) we plot the 5 most highly occupied eigenmodes of $\rho_{ij}$, which together contain some $95.4\%$ of the total field population.   
The density and phase of the most highly occupied mode [$\chi_0(\mathbf{x})$] is displayed in Fig.~\ref{fig:POsurfs_E105}(a) and contains a singly positively charged \emph{central} vortex (phase circulation of $2\pi$ around its azimuth direction).  The second most highly occupied mode [$\chi_1(\mathbf{x})$, Fig.~\ref{fig:POsurfs_E105}(b)] is a rotationally symmetric mode with uniform phase, which forms a pronounced peak at the origin.  The third most highly occupied mode [$\chi_2(\mathbf{x})$, Fig.~\ref{fig:POsurfs_E105}(c)] is another vortex mode, with vanishing central density.  However, this mode is doubly (positively) charged, with the phase varying by $4\pi$ around its azimuth.  

We note that these three modes bear a strong resemblance to a single vortex GP-eigenmode $\phi_0$ and the $u$ and $v^*$ spinor components \cite{noteI} of its so-called \emph{anomalous} Bogoliubov excitation (also referred to as the lowest core-localized state by some authors \cite{Isoshima99}).  We recall that this is the lowest-energy excitation of the single-vortex state in the Bogoliubov approximation, and its \emph{negative} energy in the laboratory frame signals the thermodynamic instability of the vortex state in that frame.  The angular-momentum numbers $m_{u,v}=(0,2)$ of the particle ($u$) and hole ($v$) functions result naturally from the transposition of the Bogoliubov pairing ansatz to the $m=1$ vortex scenario \cite{Fetter71,Dodd97}.  The collective excitation of the anomalous mode results in the displacement of the vortex off-axis in a direction determined by the phase of the Bogoliubov excitation relative to the condensate (see \cite{Rokhsar97}) in the linear expansion \cite{noteJ} 
\begin{equation}\label{eq:anomalous_mode_bog_expansion}
	\psi = a_0\phi_0 + bu + b^*v^*.
\end{equation}
In the (zero-temperature) Bogoliubov approximation the energy of the anomalous mode determines the frequency of precession of a displaced vortex under its own induced velocity field, in the appropriate linear-response limit \cite{Fetter01}.  However, this result has been shown not to hold in the finite-temperature self-consistent mean-field theories which generalize the Bogoliubov description to finite temperatures \cite{Isoshima04}.

\begin{figure}
	\includegraphics[width=0.45\textwidth]{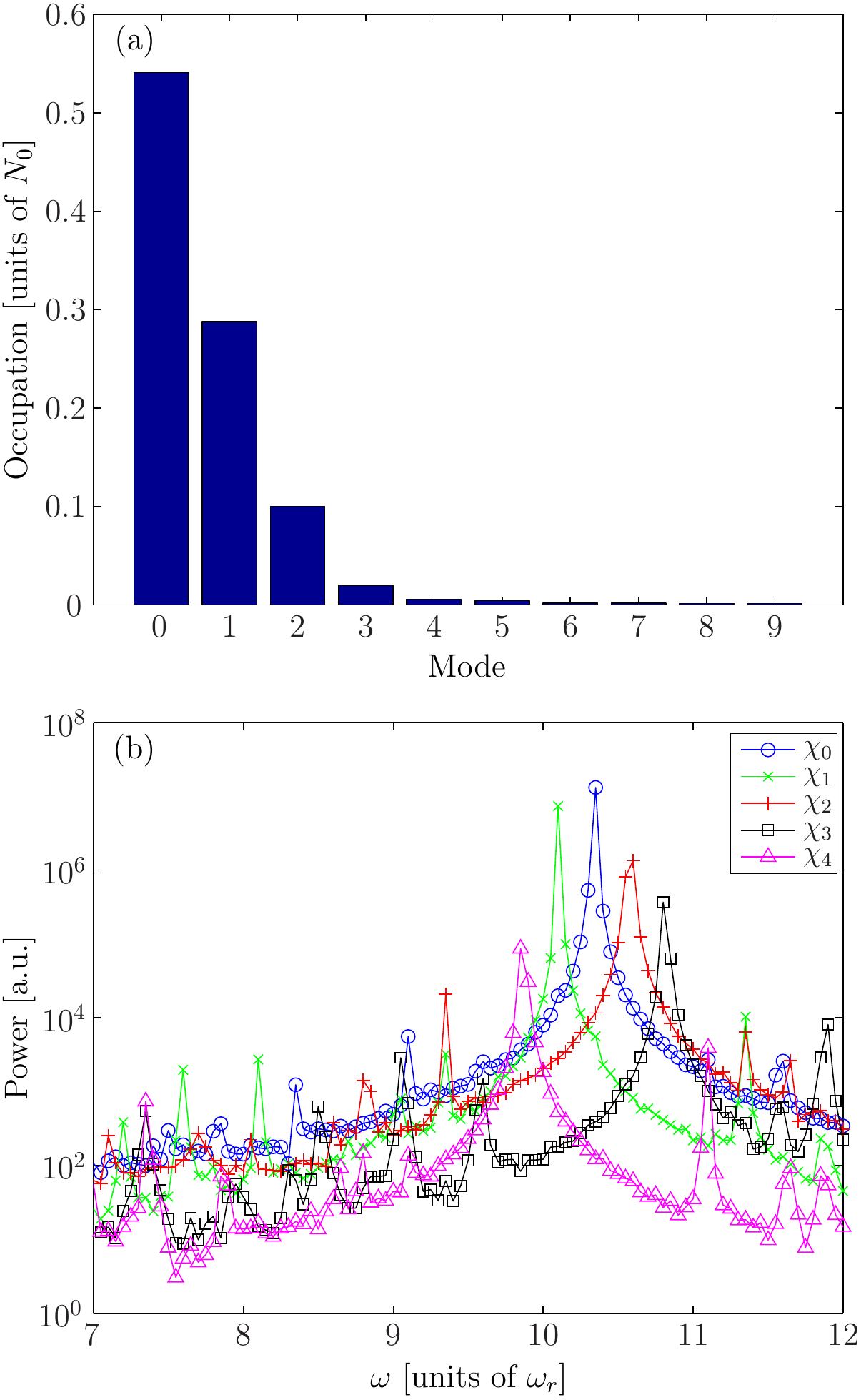}
	\caption{\label{fig:occs_spect_E105} (Color online) (a) Mean occupations of covariance matrix eigenmodes.  (b) Temporal power spectrum of coefficients $\alpha_i(t)$ of the five most highly occupied modes.  Quantities presented in each subplot are calculated for the case $E=1.05E_\mathrm{g}$.}
\end{figure}

Returning to our classical field simulation, we calculate the time-dependent expansion coefficients of the eigenmodes $\chi_i(\mathbf{x})$ over the 100 cyc.\ analysis period, defined as 
\begin{equation}\label{eq:covar_mode_coefficients}
	\alpha_i(t) \equiv \int d\mathbf{x}\; \chi_i^*(\mathbf{x})\psi(\mathbf{x},t).
\end{equation}
The average values $\langle|\alpha_i|^2\rangle_t$ of course yield simply the mode occupations $n_i$, which are plotted for the $10$ most highly occupied modes in Fig.~\ref{fig:occs_spect_E105}(a).  
From the coefficients $\{\alpha_i\}$ we form the classical correlation functions \cite{Gardiner00,Blakie05}
\begin{equation}\label{eq:define_g2}
	g_i^{(2)} = \frac{\langle|\alpha_i|^4\rangle_t}{(\langle|\alpha_i|^2\rangle_t)^2},
\end{equation}
which indicate the coherence of the modes over the analysis period.  We recall that the (local) correlation functions $g^{(n)}$ adopt values of $g^{(n)}=n!$ and $g^{(n)}=1$ for purely chaotic or thermal fields and purely coherent fields, respectively \cite{Glauber65,Blakie05}.  The distribution of $g^{(2)}$ values over occupation numbers is shown in Fig.~\ref{fig:POsurfs_E105}(f).  We find that the $g^{(2)}$ values calculated for the modes $\chi_i(\mathbf{x}) :\;0\leq i\leq 4$ are all $\lesssim1.1$.  By contrast the vast majority of the remaining modes $(i>4)$ have $g^{(2)}\approx2$.  This suggests that these $5$ most highly occupied modes are \emph{coherent}, while the majority of modes $\chi_i$ are to a large degree incoherent \emph{with respect to the averaging procedure employed here}.
In \cite{Wright08}, the temporal power spectrum of the classical field was used to analyze the coherence in the classical field.  Here we apply a similar technique to extract information from the mode coefficients $\alpha_i(t)$.  In Fig.~\ref{fig:occs_spect_E105}(b) we plot the spectra obtained from these samples, each of which was formed by averaging spectra estimates from 5 consecutive time series (see Sec.~\ref{subsec:Vortex_precession}).  
We observe that the spectrum of each mode coefficient presented in Fig.~\ref{fig:occs_spect_E105}(b) exhibits several peaks, the highest of which is in each case several orders of magnitude greater than the others.  We interpret the largest peak in each case as an indication of the quasi-uniform phase rotation associated with coherence in the classical field.  We also note that the frequency peaks of modes $\chi_1$ and $\chi_2$ are spaced evenly about the peak of $\chi_0$.  Specifically we find for the frequencies of peak power $(\omega_\mathrm{p,0},\omega_\mathrm{p,1},\omega_\mathrm{p,2})=(10.35,10.10,10.60)\omega_r$, corresponding to a quasiparticle energy of $\varepsilon_\mathrm{q}=\hbar(\omega_\mathrm{p,1}-\omega_\mathrm{p,0})\approx-0.25\hbar\omega_r$, in agreement with the vortex precession frequency $\omega_\mathrm{v}\approx0.23\omega_r$ (determined by the technique of Sec.~\ref{subsec:Vortex_precession}) to within the frequency resolution $\Delta\omega=0.05\omega_r$.   

A precessing-vortex GP state can of course be understood as a linear combination of such angular-momentum eigenmodes.  In a frame rotating at the vortex precession frequency ($\Omega_\mathrm{v}$), the phases of the various components all rotate at a single common frequency (the condensate eigenvalue $\mu/\hbar$).  In the laboratory frame these components have frequencies $\omega_m = \mu/\hbar + \Omega_\mathrm{v}m$, and their consequent dephasing over time leads to the apparent fragmentation observed in the lab frame \cite{noteK}.  Simulations of classical fields with higher energies, which contain vortices precessing at larger radii, when subjected to the above analysis, again yield a decomposition into angular-momentum eigenmodes.  However, in such cases we find that the most highly occupied mode has $m=0$, as the anomalous particle component $u$ has grown, pushing the vortex further from the trap center  and replacing the $m=1$ vortex mode as most highly occupied.
%%%%%%%%%%%%%%%%%%%%%%%%%%%%%%%%%%%%%%%%%%%%%%%%%%%%%%%%%%%
\subsubsection{Rotating frames}\label{subsec:PO_rotating_frames}
\begin{figure*}
	\includegraphics[width=0.9\textwidth]{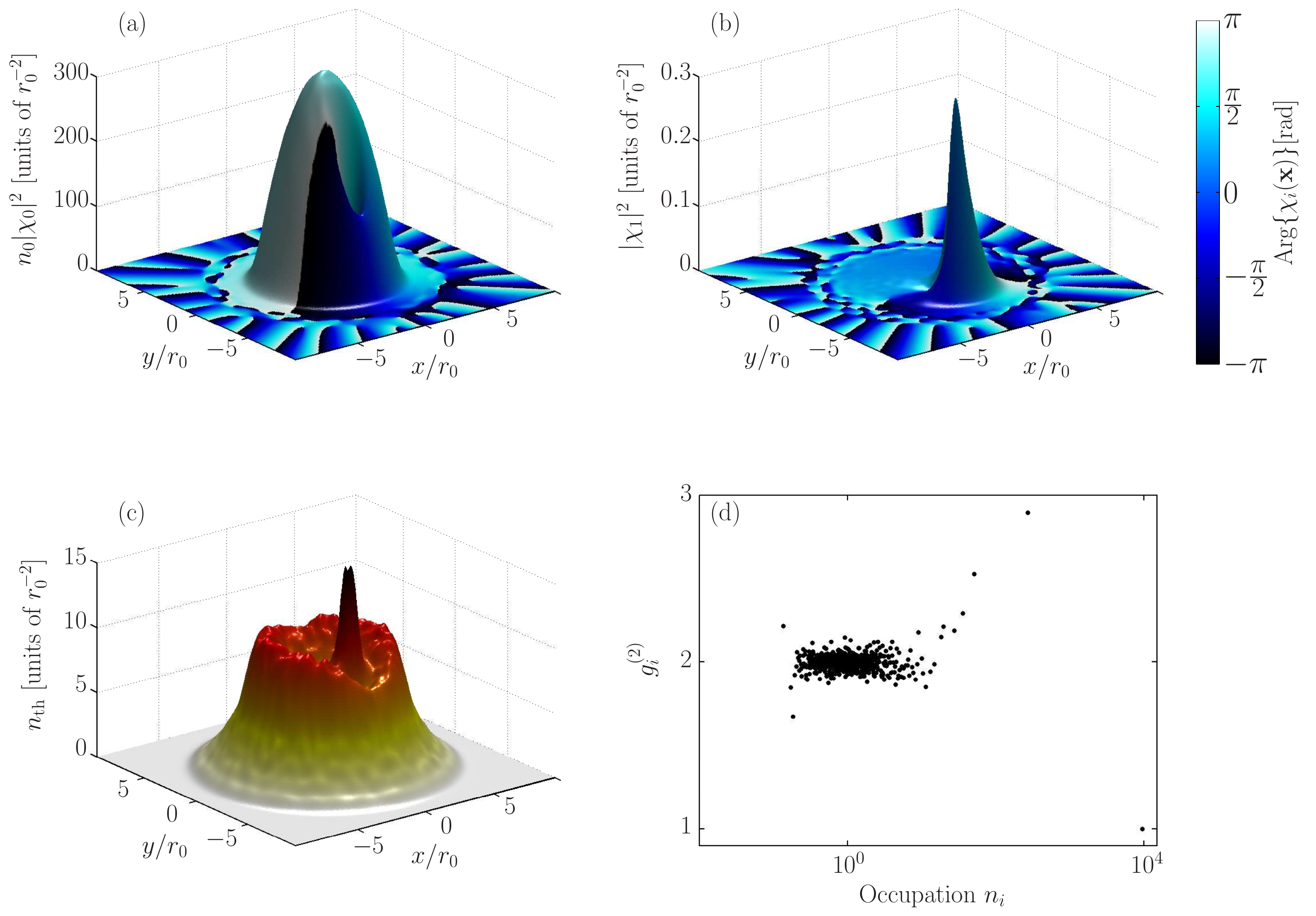}
	\caption{\label{fig:POsurfs_E110_corot} (Color online) Density and phase of (a) the condensate and (b) the Goldstone mode, as determined from the covariance matrix formed in the co-rotating frame.  The condensate density is the density of mode $\chi_0(\mathbf{x})$, shown normalized to its occupation $n_0$. (c) Thermal density of the field. (d) Second-order coherence functions of the covariance-matrix eigenmodes versus mode occupation.} 
\end{figure*}

The above results of a naive application of the PO procedure in the laboratory frame suggest that the classical field condenses in a rotating frame.  The mean phase rotation of the various angular-momentum components is such that upon transformation to an appropriate rotating frame these modes collectively exhibit quasi-uniform phase rotation at a common frequency.  We therefore expect that transforming the field to its representation in such a rotating frame, before applying the PO procedure, will yield a single condensate mode.  We consider now a simulation with $E=1.10E_\mathrm{g}$, where in equilibrium a single vortex precesses with mean displacement of $\overline{r_\mathrm{v}}\approx2.0r_0$ from the trap center.  After transforming the field coefficients $c_{nl}$ to a frame (the \emph{measurement} frame) with fixed rotation frequency $\Omega_\mathrm{m}$, we construct the covariance matrix in the same manner as described in Sec.~\ref{subsubsec:lab-frame_PO}.  In Fig.~\ref{fig:POsurfs_E110_corot}(a-b) we plot the two most highly occupied modes obtained from this covariance matrix, having performed the PO procedure in a frame with $\Omega_\mathrm{m}=0.2610\omega_r$.  We discuss below the reasons for this choice of rotation frequency.  
The most highly occupied mode is now a vortical state in which the vortex is \emph{displaced} off-axis, similarly to the case for the full classical field.  The second most highly occupied mode is now a peaked function which partially `fills' the density dip associated with the vortex in $\chi_0$.  The prominent peak of this mode exhibits a distinctive amplitude and phase pattern, which we identify as that of the \emph{Goldstone mode} associated with the breaking of rotational symmetry by the off-axis vortex state.  Such a mode arises in scenarios in which the condensate mode $\phi_0$ breaks a symmetry of the system Hamiltonian.  In the case of broken rotational symmetry \cite{Lobo05}, the Goldstone mode has the form $(u_\mathrm{G},v_\mathrm{G})\sim(L_z \phi_0,-L_z^*\phi_0^*)$, and corresponds to small rotations of the symmetry-broken condensate mode.  As such a Goldstone mode is of a fundamentally different nature to the `proper' normal modes (quasiparticles) of the system \cite{Blaizot85}, we regard this mode as separate from the thermal or noncondensate fraction of the field \cite{noteU}.  We therefore form the thermal density as the total density of all modes other than the condensate mode $(\chi_0)$ and Goldstone mode $(\chi_1)$,
\begin{equation}\label{eq:thermal_density}
	n_\mathrm{th}(\mathbf{x}) = \sum_{i\geq2} n_i |\chi_i(\mathbf{x})|^2.
\end{equation}
We plot this density in Fig.~\ref{fig:POsurfs_E110_corot}(c).  We note that the thermal density exhibits a sharp peak at the vortex location, resulting from the erratic motion of the vortex about its mean trajectory in the classical-field evolution.  Thus a careful application of the PO procedure yields a core-filling thermal component, as is predicted by the Hartree-Fock-Bogoliubov family of self-consistent mean-field theories \cite{Virtanen01}.  This is in contrast to the instantaneous classical field, which, due to its essential nature as a single-particle wavefunction (i.e., vector in the projected single-particle Hilbert space) contains only `bare' vortices, with no possibility of core filling.  In the microcanonical-ergodic approach of the PGPE \cite{Blakie05,Blakie08} the two (and higher) particle correlations emerge from the field when averaged over time.  However, an alternative interpretation of classical-field trajectories (such as those of \cite{Wright08}) as representative of single experimental trajectories can be made, but, as noted in \cite{Blakie08}, this interpretation must be made with caution.  In particular it is \emph{not correct} to interpret the instantaneous state of the many-body wavefunction as the Hartree product of the classical field $\psi(\mathbf{x})$ (as is the interpretation of the GP wavefunction in the zero-temperature GP theory \cite{Leggett01,noteL}.)  Rather, the field $\psi(\mathbf{x})$ must be interpreted as an approximation to the quantum \emph{field operator}, from which certain correlations can be deduced.  Some averaging of this instantaneous field $\psi(\mathbf{x})$, over an ensemble of trajectories \cite{Norrie06a} or over time \cite{Blakie05} must be performed to construct the correlations necessary to describe the thermal core-filling of a vortex. 

\begin{figure}
	\includegraphics[width=0.45\textwidth]{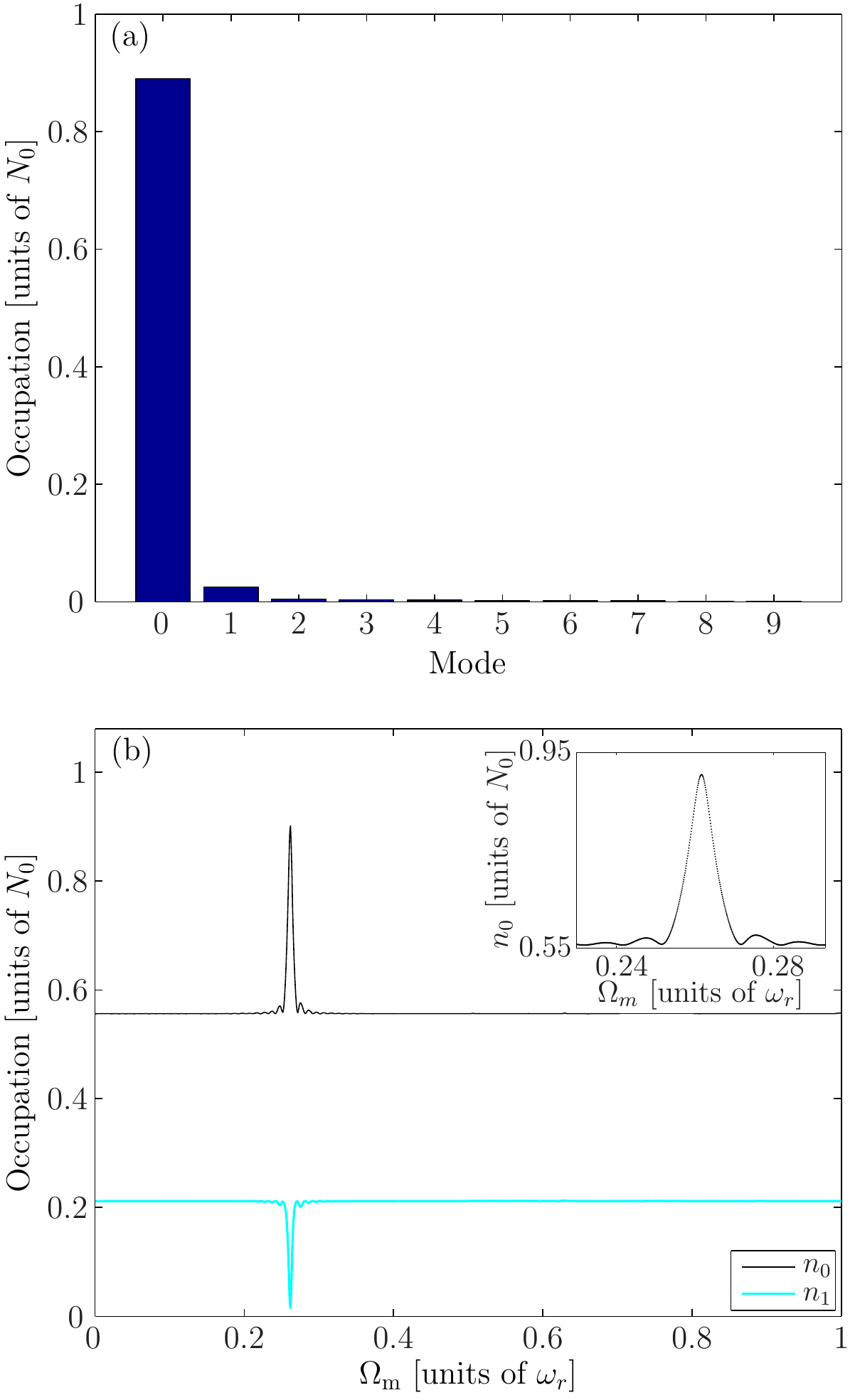}
	\caption{\label{fig:fracs_and_occs_corot} (Color online) (a) Occupations of the 10 most highly occupied eigenmodes of the covariance matrix determined in the frame with $\Omega_\mathrm{m}=0.2610\omega_r$. (b) Dependence of mode populations on the rotation frequency of the frame in which the covariance matrix is constructed. The inset shows the behavior of $n_0$ close to its maximum.}
\end{figure}

We calculate the time-dependent expansion coefficients of the modes $\chi_i(\mathbf{x})$ in the same manner as described in Sec.~\ref{subsubsec:lab-frame_PO}.  The mean occupations of the modes are plotted in Fig.~\ref{fig:fracs_and_occs_corot}(a), and reveal that a single mode contains 89.1\% of the field population.  We present in Fig.~\ref{fig:POsurfs_E110_corot}(d) the coherence functions obtained from the mode coefficients.  The most highly occupied mode is now the only mode with $g^{(2)}$ significantly less than the thermal value $g^{(2)}=2$.  We find in fact $g^{(2)}_0 = 1.0002$, indicating very high coherence of the condensate mode over the averaging period.  The next most highly occupied modes now have values $g^{(2)}\gtrsim2.2$, \emph{greater} than that expected for thermally or incoherently occupied single-particle modes.  This is a signature of the strongly \emph{collective} nature of the low-lying excitations of the condensate.  The form of these excitations is strongly affected by the presence of the interacting, macroscopically occupied condensate, which induces anomalous (pairing) correlations in the field, which in turn manifest as anomalous values for the coherence functions \cite{Gardiner01}.  We note that the Goldstone mode has $g^{(2)}=2.90$, close to the upper bound $g^{(2)}=3$ \cite{Gardiner01}, as we might expect given that a Goldstone mode, with $u_\mathrm{G}=-v_\mathrm{G}^*$, is the most strongly collective excitation possible.  The fact that these pair correlations are manifest in the density matrix formed in the co-rotating frame is further evidence that this frame gives the correct resolution of the condensate.  It is interesting to note that the anomalous nature of these correlations is not apparent in coordinate-space representations of the field \cite{Blakie05}, or in the lab-frame analysis of Sec.~\ref{subsubsec:lab-frame_PO}.  A more natural characterization of the low-lying excitations would exploit these anomalous one-body correlations in order to construct an appropriate quasiparticle basis, this subject will be discussed elsewhere \cite{noteM}.

The choice of measurement-frame frequency ($\Omega_\mathrm{m}=0.2610\omega_r$) optimizes the resolution of the condensate as we now discuss.  From the discussion of Sec.~\ref{subsubsec:lab-frame_PO}, we expect (neglecting for now the role of the Goldstone mode) that the classical field can be resolved into a single condensate mode coexisting with a bath of thermal atoms, in some rotating frame.  In general the rotation frequency of the measurement frame differs from the equilibrium rotation frequency of the condensate, and this difference leads to dephasing of the distinct angular-momentum components of the condensate mode, and hence to a spurious fragmentation of the condensate into a set of temporally coherent angular-momentum eigenmodes.  We propose as a model for analysis of classical-field data, that in the frame in which condensation occurs, the field should decompose into a single condensate mode plus thermal material upon performing the PO procedure.  This frame depends on a single parameter (the measurement-frame rotation frequency $\Omega_\mathrm{m}$), and the magnitude of the largest eigenvector, regarded as a function of the measurement-frame frequency [$n_0=n_0(\Omega_\mathrm{m})$] forms the corresponding \emph{optimality function}.  Our procedure then is to maximize the optimality function to determine the optimal measurement-frame frequency $\Omega_\mathrm{c}$, which is the frequency of the rotating frame in which the condensate mode is (most) stationary.  This provides a sensitive measure of the (lab-frame) rotation frequency of the vortical condensate mode, and provides a unified basis for the calculation of the properties of the condensed and noncondensed material in the classical field.  In particular, we will regard the frame of vortex precession and the frame of condensation as interchangeable, though in practice we will use the latter exclusively in the remainder of this paper.
In Fig.~\ref{fig:fracs_and_occs_corot}(b) we present results of this optimization procedure as applied to our simulation with $E=1.10E_\mathrm{g}$ over a $100$ cyc.\ period starting from $t=9000$ cyc.  We observe a prominent peak in $n_0(\Omega_\mathrm{m})$ at $\Omega_\mathrm{m}=0.2610\omega_r$.  We also plot the magnitude of the second largest eigenvalue $n_1(\Omega_\mathrm{m})$, and observe that it exhibits a dip at this same frequency, indicating the sensitivity of the one-body correlations to the frequency $\Omega_\mathrm{m}$.  We take the location of the peak in $n_0(\Omega_\mathrm{m})$ as an estimate for the rotation frequency $\Omega_\mathrm{c}$ of the condensate \cite{noteN}.   
%%%%%%%%%%%%%%%%%%%%%%%%%%%%%%%%%%%%%%%%%%%%%%%%%%%%%%%%%%%
\subsubsection{Temporal decoherence and sample length}\label{subsubsec:temporal_decoherence}
\begin{figure}
	\includegraphics[width=0.45\textwidth]{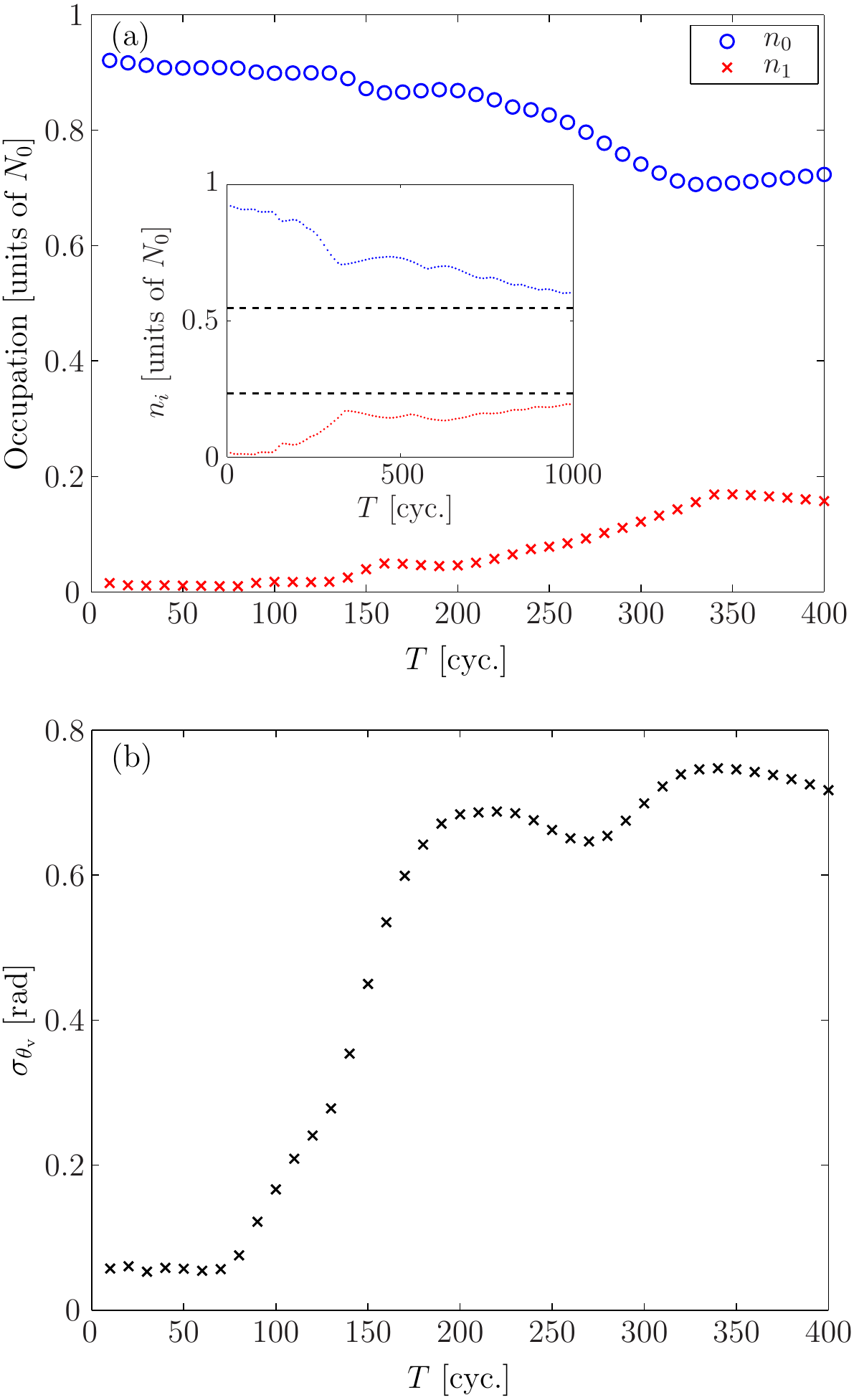}
	\caption{\label{fig:temporal_decay} (Color online) (a) Largest eigenvalues ($n_0$ and $n_1$) of the covariance matrix as a function of averaging time. Inset shows the behavior of $n_0$ and $n_1$ (upper and lower dotted lines, respectively) over longer times, where the eigenvalues approach their values obtained in the laboratory frame (dashed lines). (b) Standard deviation of vortex phases measured in a frame rotating at $\Omega=0.2610\omega_r$.}
\end{figure}

We now consider how the condensate mode determined by the above procedure decays as a function of the time over which the averaging procedure is performed.  As already noted in Sec.~\ref{subsec:Vortex_precession}, the phase of the vortex (with respect to any uniform rotation) diffuses over time.  This diffusion is intimately related to the breaking of the rotational symmetry of the system by the presence of the off-axis vortex, as we now discuss.  The condensate orbital formed in the classical field is an off-axis vortex mode, which is not an eigenstate of angular momentum and thus not rotationally invariant.  This breaks the rotational [$\mathrm{SO}(2)$] symmetry of the Hamiltonian Eq.~(\ref{eq:HCF}) \cite{noteP}.  A well-known consequence of the breaking of a symmetry by a dynamical equilibrium in classical mechanics is the appearance of a zero-energy normal mode \cite{Goldstein50}, and this result persists in the quantum theory, where broken symmetries are similarly accompanied by so-called spurious or Goldstone modes \cite{Blaizot85}. 
Such excitations are associated with a collective motion without restoring force \cite{Blaizot85,Lewenstein96}.  In the context of classical-field theory, we expect that the erratic evolution of the field results in random excitation of this motion, corresponding to fluctuations of the vortex phase.  As there is no `restoring force' associated with this excitation, over time the phase can drift arbitrarily far from any initial value, invalidating the interpretation of the equilibrium as a configuration in which energy is equipartitioned over normal excitations of the symmetry-broken condensate orbital.  As a consequence, the decomposition of the field into condensate and noncondensate by the Penrose-Onsager procedure also deteriorates over time.  This is of course consistent with our expectation that as the averaging time becomes large, the statistics we measure converge to those of the ergodic density of the microcanonical system, which must reflect the rotational symmetry of the Hamiltonian \cite{Lebowitz73}.
In Fig.~\ref{fig:temporal_decay}(a) we plot the values of the two largest eigenvalues of the covariance matrix formed by averaging for varying lengths of time, each average starting at $t=9000$ cycles.  For each averaging time we have optimized the condensate fraction over the frequency of frame rotation, which causes only small variation in the optimal frequency ($\lesssim0.003\hbar\omega_r$).  We observe that the largest eigenvalue gets smaller as the averaging period increases, while the second largest eigenvalue gets larger.  Over longer periods [inset to Fig.~\ref{fig:temporal_decay}(a)] the two eigenvalues approach the values obtained from the density matrix constructed in the laboratory frame (dashed lines), indicating that the orientation of the condensate mode has diffused to such an extent that the density matrix describes a fragmented condensate in all uniformly rotating frames.  In Fig.~\ref{fig:temporal_decay}(b) we plot the standard deviation of the vortex phases as a function of time, in a frame rotating at $\Omega=0.2610\omega_r$.  We observe that for the first $\approx75$ trap cycles the standard deviation remains reasonably steady at a small value $\sigma_{\theta_\mathrm{v}}\sim0.05$ radians.  After this the standard deviation begins to increase dramatically, and this is ultimately reflected in the measured eigenvalues [Fig.~\ref{fig:temporal_decay}(a)].  We note that the standard deviation we have presented is simply that of the distribution of vortex phases measured in a single trajectory, as a function of the length of time over which vortex phases are accumulated.  To properly characterize the rate of rotational diffusion would require the calculation of the variance of the vortex-phase change over an ensemble of similarly prepared classical-field trajectories (c.f. \cite{Sinatra08}), a numerically heavy task which we do not pursue here.  We note further that the initial period of comparative stability of the vortex phase is not a generic feature of the system, in general the excursions of the vortex phase relative to its mean precessional motion occur unpredictably.  We note however that the condensate fraction obtained by averaging for only 10 cyc.\ already agrees with the values throughout this period of stability to within $2-3\%$, in agreement with the findings of \cite{Blakie05}.  We therefore adopt the following approach to determining the condensate in the remainder of this paper: we form the covariance matrix $\rho_{ij}$ from samples over a period of $T=10$ cyc., in a frame rotating at frequency $\Omega_\mathrm{m}$.  We vary this frequency so that the largest eigenvalue of the corresponding covariance matrix is maximized.  We take the frequency of this optimal frame as an estimate of the angular velocity of the condensate, and the largest eigenvalue as an estimate of the condensate fraction, and similarly for subsidiary quantities obtained from the PO decomposition (see Sec.~\ref{subsec:rotational_properties}).  We do this for 100 consecutive periods of 10 cycles, over the period $t=9000-10000$ cyc.\ and average the estimates for $\Omega_\mathrm{m}$ and $f_\mathrm{c}$.  In this way we sample the short-time dynamics which determine the correlations of interest, while exploiting the ergodic character of the classical-field evolution over longer times.  For the simulation with $E=1.10E_\mathrm{g}$, we find from this procedure $\Omega_\mathrm{c}=0.2598\omega_r$ and $f_\mathrm{c}=0.9110$, in reasonably close agreement with the values obtained from a single averaging over a longer time period.
We conclude this discussion by considering the interpretation of the condensate fraction we measure, in light of our abandonment of formal ergodic averaging.  We interpret the results of the short-time averaging procedure in terms of a symmetry-broken representation \cite{Castin04} of the many-body system.  The one-body density matrix we calculate corresponds to a particular (mean) orientation of the condensate mode (vortex phase $\theta_\mathrm{v}$).  Labeling this density matrix by $\rho_{\theta_\mathrm{v}}$, the full one-body density matrix is obtained by integrating over all possible vortex phases $\rho^{(1)} = (1/N)\int d\theta_\mathrm{v} \;  \rho_{\theta_\mathrm{v}}^{(1)}$ \cite{Castin04,Seiringer08}, with $N$ a normalization factor.  This one-body density matrix does not exhibit condensation in the Penrose-Onsager sense, and this is true of the physical $N$-body system also.  Nevertheless, the individual one-body density matrix $\rho_{\theta_\mathrm{v}}$ exhibits a condensate, and we interpret the correlations it describes as characterizing the behavior of a corresponding physical system \emph{observed} to have a particular vortex orientation.  With this in mind, we will continue to refer to the largest eigenvalue of the density matrix $\rho_{\theta_\mathrm{v}}$ constructed from the classical field trajectory as the condensate fraction.
%%%%%%%%%%%%%%%%%%%%%%%%%%%%%%%%%%%%%%%%%%%%%%%%%%%%%%%%%%%
%%%%%%%%%%%%%%%%%%%%%%%%%%%%%%%%%%%%%%%%%%%%%%%%%%%%%%%%%%%%%%%%%%%%%%%%%%%%%%%%%%%%%%%%%%%%%%%%%%%
\subsection{Rotational properties of the field}\label{subsec:rotational_properties}
%%%%%%%%%%%%%%%%%%%%%%%%%%%%%%%%%%%%%%%%%%%%%%%%%%%%%%%%%%%%%%%%%%%%%%%%%%%%%%%%%%%%%%%%%%%%%%%%%%%
The prescription we have developed for decomposing the classical field into condensed and noncondensed parts allows us to extract quantities which characterize the rotation of the two components individually, by calculating appropriate expectation values.  We recall that the extension of a single-body operator $J$ to a second-quantized operator $\hat{J}=\int d\mathbf{r} \hat{\psi}^\dagger J \hat{\psi}$ on the many-particle Fock space \cite{Fetter71a} has expectation value in the many-body state $\langle\hat{J}\rangle = \mathrm{Tr}\{\hat{\rho}^{(1)}J\}$.  We thus define an expectation value of a single-body operator in the classical field analogously by 
\begin{equation}
	\langle J \rangle_\mathrm{c} \equiv  \mathrm{Tr}\{\rho J\},
\end{equation}
where $\rho$ is the covariance matrix introduced in Sec.~\ref{subsec:Penrose_Onsager_analysis}.
The decomposition into a condensed mode and noncondensed modes introduced in Sec.~\ref{subsec:PO_rotating_frames} \cite{noteW} allows us to write (using Dirac notation for vectors in the projected single-particle space)
\begin{equation}
	\rho = n_0|\chi_0\rangle\langle \chi_0| + \sum_{k>0} n_k |\chi_k\rangle\langle \chi_k| \equiv \rho_0 + \rho_\mathrm{th}.
\end{equation}
We thus define averages in the condensate and noncondensate by 
\begin{equation}
	\langle J\rangle_0 = \mathrm{Tr}\{\rho_0J\} = n_0\sum_m\langle m|\chi_0\rangle\langle \chi_0|J|m\rangle,
\end{equation}
and
\begin{equation} 
	\langle J\rangle_\mathrm{th} = \mathrm{Tr}\{\rho_\mathrm{th}J\} = \sum_{k>0}n_k\sum_m\langle m|\chi_k\rangle\langle \chi_k|J|m\rangle,
\end{equation}
respectively.
It is clear from the form of $\rho_0$ and $\rho_\mathrm{th}$ that they are (proportional to) mutually orthogonal projectors, and so we have the additivity property $\langle J\rangle_0 + \langle J\rangle_\mathrm{th} = \langle J \rangle_\mathrm{c}$.
%%%%%%%%%%%%%%%%%%%%%%%%%%%%%%%%%%%%%%%%%%%%%%%%%%%%%%%%%%%
\subsubsection{Angular momentum of the condensate}
As noted in Sec.~\ref{subsec:PO_rotating_frames}, in the precessing-vortex scenario the condensate determined by our method is a non-axisymmetric state with an off-center vortex.  Analogous condensate orbitals appear in the zero-temperature GP theory as intermediates between the rotationally invariant $\ell=0$ and $\ell=1$ modes, with fractional angular-momentum (per atom) expectation values $0 < \ell < 1$ \cite{Butts99,Papanicolaou05}, and are mechanically unstable (see the discussion in \cite{Komineas05}).  

Applying the above averaging procedure to our simulation with $E=1.10E_\mathrm{g}$ we obtain $L_\mathrm{c} \equiv \langle L_z\rangle_0 / n_0 = 0.63\hbar$.  We recall that the condensate fraction obtained for this state was $f_\mathrm{c}=0.911$.  Turning to the noncondensate we find $L_\mathrm{th}/N_\mathrm{th} = 4.80\hbar$.  While at this energy, only $9\%$ of the atoms have been excited out of the condensate, in order to maintain rotational equilibrium of the system, the thermal atoms carry nearly half of the angular momentum of the field.  This is a consequence of the suppressed moment of inertia of the condensate: in order to have the same angular velocity, the thermal cloud must possess much more angular momentum per particle than the condensate \cite{Guery-Odelin00,Haljan01a}.
%%%%%%%%%%%%%%%%%%%%%%%%%%%%%%%%%%%%%%%%%%%%%%%%%%%%%%%%%%%
\subsubsection{Angular velocity of the thermal cloud}\label{subsubsec:cloud_rotation}
Now we consider the angular velocity of the thermal cloud.  In \cite{Wright08} we gave an approximate method of separating the rotational properties of the condensate and the thermal cloud, based entirely on spatial location.  In the current paper, where we have developed a rigorous procedure for separating the condensate and thermal components, we can give a more accurate analysis of the rotation and inertia of both of these two components.
In particular our approach includes the contribution of thermal fluctuations that traverse the central condensed region, as well as those which fill the vortex core, to the total thermal component of the field.  We thus calculate the averages $\langle L_z\rangle_\mathrm{th}$ and $\langle \Theta_\mathrm{c} \rangle_\mathrm{th}$, and estimate the cloud rotation frequency by 
\begin{equation}
	\Omega_\mathrm{th} \approx \frac{\langle L_z \rangle_\mathrm{th}}{\langle \Theta_\mathrm{c} \rangle_\mathrm{th}},
\end{equation}
i.e., we \emph{assume} that the thermal component's moment of inertia is equal to its classical value. Doing so we find $\Omega_\mathrm{th} = 0.262\omega_r$,  in fair agreement with the precession frequency of the vortex.  
%%%%%%%%%%%%%%%%%%%%%%%%%%%%%%%%%%%%%%%%%%%%%%%%%%%%%%%%%%%%%%%%%%%%%%%%%%%%%%%%%%%%%%%%%%%%%%%%%%%%%%%%%%%%%%%%%%%%%%%%%%%%%%%%%%%%
\section{Dependence of system state on internal energy}\label{sec:Energy_dependence}
We now consider the effect of varying the energy of the classical field on its equilibrium properties.  We expect quite generically for Hamiltonian classical-field simulations such as those performed here that an increase in the classical-field energy will result in an increase in the field temperature and a suppression of the condensate fraction \cite{Blakie05}.  In this scenario with finite conserved angular momentum, we also expect the rotational properties of the system to depend strongly on the field energy.
%%%%%%%%%%%%%%%%%%%%%%%%%%%%%%%%%%%%%%%%%%%%%%%%%%%%%%%%%%%
\subsection{Density}
\begin{figure*}
	\includegraphics[width=0.9\textwidth]{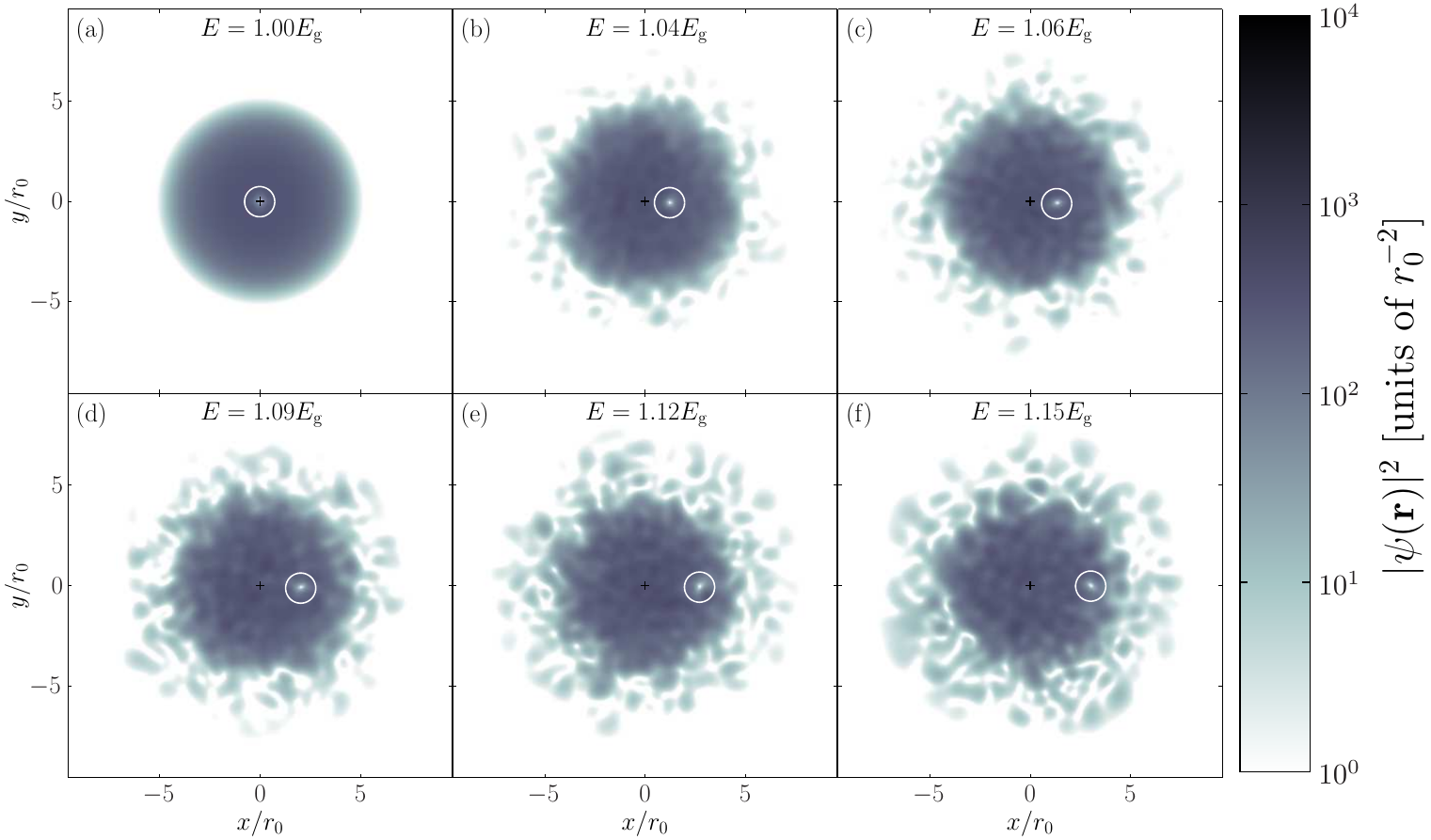}
	\caption{\label{fig:density_vary_E} (Color online) (a-f) Representative densities of equilibrium classical fields of different energies. The white circle indicates the vortex position, and $+$ marks the coordinate origin (trap axis).} 
\end{figure*}

The most obvious and generic consequence of increasing the energy of the microcanonical system is an increase in the entropy of the system.  The qualitative effects of this increase can be readily observed in position-space representations of the classical field density.  In Fig.~\ref{fig:density_vary_E}(a-f) we plot representative densities of the classical field at equilibrium, for various values of the field energy.  
These densities are chosen at times in the range $t\in[9000,10000]$ cyc.\ such that the vortex is displaced along the $x$ axis, for ease of comparison. We clearly observe from these images an increase in the surface excitations of the condensate and the development of a turbulent outer cloud as the energy is increased.  The images also reveal an increase in the vortex precession radius as the energy is increased.   
%%%%%%%%%%%%%%%%%%%%%%%%%%%%%%%%%%%%%%%%%%%%%%%%%%%%%%%%%%%
\subsection{Condensate fraction and angular momentum}\label{subsec:cfrac_and_angmom}
From the procedure introduced in Sec.~\ref{subsubsec:temporal_decoherence}, we find the condensate fraction [Fig.~\ref{fig:fns_of_energy}(a)] drops gradually from $f_\mathrm{c}=0.96$ to $f_\mathrm{c}=0.83$ over the energy range $E\in[1.04,1.19]E_\mathrm{g}$.  However, the angular momentum (per particle) of the condensate drops dramatically, falling from $\langle L_z\rangle_0/n_0$ = $0.86\hbar$ at $E=1.04E_\mathrm{g}$ to nearly zero by $E=1.16E_\mathrm{g}$.  This is a signature of the anomalous rotational properties of the off-axis vortex state \cite{Papanicolaou05}: as the energy is increased the condensate loses angular momentum to the cloud and the rotation rate of \emph{both} components increases.  

At higher energies the condensate is vortex-free and thus irrotational, and exists in thermal and diffusive equilibrium with the thermal cloud which now contains all the angular momentum of the field.  The small residual angular momentum of these high-energy condensates shown in Fig.~\ref{fig:fns_of_energy}(a) results from surface excitations which are not differentiated from the condensate by the short-time averaging.  We note however that the condensate fraction is essentially continuous across this transition from a symmetry-broken condensate orbital to a non-symmetry-broken (i.e., vortex-free) one, supporting the definition introduced in Sec.~\ref{subsubsec:temporal_decoherence} as an appropriate measure of condensation in the symmetry-broken regime.
%%%%%%%%%%%%%%%%%%%%%%%%%%%%%%%%%%%%%%%%%%%%%%%%%%%%%%%%%%%
\subsection{Condensate and cloud rotation rates}\label{subsec:rotation_rates}
In Fig.~\ref{fig:fns_of_energy}(b) we plot the condensate rotation frequency, as determined by the frame frequency of optimal condensate resolution (Sec.~\ref{subsec:PO_rotating_frames}).  We find that the rotation frequency increases gradually from $\Omega_\mathrm{c}=0.234\omega_r$ at $E=1.04E_\mathrm{g}$, up to $\Omega_\mathrm{c}=0.340\omega_r$ at $E=1.15E_\mathrm{g}$.  For comparison the frequency of the anomalous mode of the ground vortex state in the Bogoliubov approximation, $|\epsilon_\mathrm{anom}|/\hbar=0.2034\omega_r$, is plotted as a dotted line in Fig.~\ref{fig:fns_of_energy}(b).  We note that the case $E=1.15E_\mathrm{g}$ is very close to the transition from a vortical condensate to a vortex-free one; in $\sim10\%$ of the 10 cyc.\ sampling periods the condensate is vortex-free, resulting in maximal values of $n_0(\Omega_\mathrm{m})$ occurring around $\Omega_\mathrm{m}=1\omega_r$ \cite{noteT}.  We discard these values in calculating the mean and standard deviation presented for this simulation in Fig.~\ref{fig:fns_of_energy}(b).   From Fig.~\ref{fig:fns_of_energy}(a) we observe that the condensate modes with $E\geq1.16E_\mathrm{g}$ are essentially irrotational, and indeed the condensates in these simulations appear to be largely vortex-free, the majority of samples yielding $\Omega_\mathrm{c}\approx1\omega_r$, with a small minority producing $\Omega_\mathrm{c}\sim0.4\omega_r$, resulting from the transit of short-wavelength surface excitations (so-called \emph{ghost} vortices) which produce a slightly greater calculated condensate fraction in a frame which matches their motion, and we omit these points from Fig.~\ref{fig:fns_of_energy}(b).    
 
Turning our attention to the cloud rotation frequency [crosses in Fig.~\ref{fig:fns_of_energy}(b) (red online)], determined by the prescription of Sec.~\ref{subsubsec:cloud_rotation},  we find these results in qualitative agreement with the measured condensate rotation frequency over the range $E\in[1.04,1.15]E_\mathrm{g}$.  The bars accompanying the markers for condensate and cloud rotation rates simply represent the statistical dispersion in the measured quantities as discussed in Sec.~\ref{subsubsec:temporal_decoherence}, and \emph{not} quantitative measures of the errors in these measurements, but it is clear that the measured cloud rotation rate does not match the condensate rotation rate closely.  It is possible that this is associated with pair correlations in the noncondensate, though if anything we expect the presence of pair correlations to suppress the moment of inertia relative to its classical value (c.f. \cite{Sorensen73}).  We note finally that the cloud frequency reaches a maximum of $\Omega_\mathrm{th}\approx0.33\omega_r$ at $E=1.16E_\mathrm{g}$ before decreasing as the energy is increased further.  This reduction of $\Omega_\mathrm{th}$ with increasing energy is easily understood:  once the condensate becomes vortex-free (irrotational), as further atoms are depleted from the condensate the thermal cloud population increases while its angular momentum remains fixed, implying a decrease in angular velocity.
%%%%%%%%%%%%%%%%%%%%%%%%%%%%%%%%%%%%%%%%%%%%%%%%%%%%%%%%%%%
\subsection{Temperature and chemical potential}
\begin{figure*}
	\includegraphics[width=0.9\textwidth]{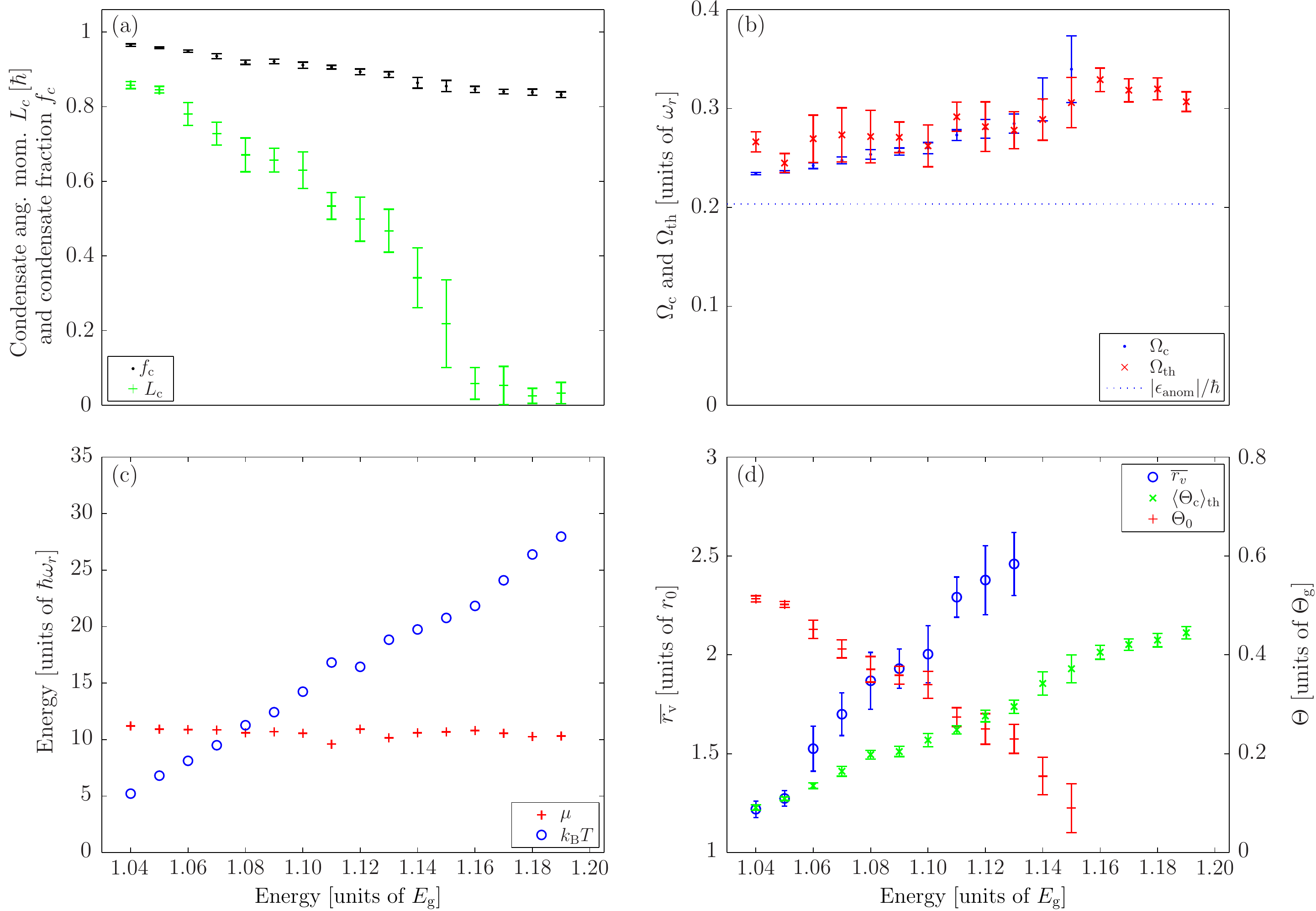}
	\caption{\label{fig:fns_of_energy} (Color online) Dependence of equilibrium parameters on field energy. (a) Condensate fraction and condensate angular momentum. (b) Angular velocities of the condensate and thermal cloud. (c) Temperature and chemical potential. (d) Vortex precession radius, classical moment of inertia of the thermal cloud, and nonclassical moment of inertia of the condensate mode.  Errorbars indicate the statistical spread in values calculated (see main text).}
\end{figure*}

In order to estimate the temperature and chemical potential of the field, we fit the distribution of thermal atoms in the classical field with an approximate semiclassical distribution \cite{Wright08}.  The principle of the method is to form a fitting function corresponding to the semiclassical distribution of atoms in the combined potential formed from the centrifugally dilated trapping potential and the mean-field repulsion of the (time-averaged) classical field.  In contrast to \cite{Wright08}, the rotation rate of the thermal cloud, which defines the centrifugal dilation of the trapping potential experienced by the thermal atoms, is not assumed to be the same as that of the frame in which the projector is applied.  We are therefore obliged to generalize the fitting function of \cite{Wright08} to allow for a differential rotation between the thermal cloud and projector.  The details of this generalization are included in Appendix~\ref{app:AppendixB}, where we derive the form of the fitting function
\begin{equation}\label{eq:fitting_function}
n(r;\mu,T) = \frac{1}{\lambda_\mathrm{dB}^2} \Big[I_1(r;\mu,\Omega_\mathrm{th},\Omega_\mathrm{p}) + I_2(r;\mu,\Omega_\mathrm{th},\Omega_\mathrm{p})\Big].
\end{equation}
In order to avoid including significant condensate density in our fitting, we are obliged to restrict our fit to a domain $r\in[r_-,r_\mathrm{tp}]$, where $r_-$ is the location of the minimum of the effective potential, which we expect to approximately mark the condensate boundary, and $r_\mathrm{tp}=\sqrt{2E_R/m(\omega_r^2-\Omega_\mathrm{p}^2)}$ is the semiclassical turning point of the condensate band in the frame of the projector \cite{Wright08}.  In practice the low temperatures of the classical-field configurations we consider make performing such a fit difficult, in particular the fit fails close to $r_-$, in the region of strongly collective excitations of the condensate.  We therefore perform our fits only to the wing of the classical-field density; for definiteness we take $r\in[r_-+1r_0,r_\mathrm{tp}]$.  We take the values $\Omega_\mathrm{th}$ calculated in Sec.~\ref{subsec:rotation_rates} as estimates for the cloud rotation rate in Eq.~(\ref{eq:fitting_function}).  Although these values may be somewhat inaccurate as discussed in Sec.~\ref{subsec:rotation_rates}, we find that using the vortex precession frequencies (where available) makes little difference to the obtained values of $\mu$ and $T$, and conclude that any discrepancy in $\Omega_\mathrm{th}$ is inconsequential at the level of accuracy of the semiclassical estimates we seek.  The results of this fitting procedure are presented in Fig.~\ref{fig:fns_of_energy}(c).
We observe that the temperature estimate $k_\mathrm{B}T$ increases approximately linearly with the energy $E$, and is of the same order of magnitude as the chemical potential $\mu$.  That the chemical-potential estimates are generally higher than the chemical potential of the ground state $\mu_\mathrm{g}=10.35\hbar\omega_r$, whereas we expect the chemical potential of the strictly Hamiltonian system to decrease with increasing energy (see \cite{Davis05}) suggests that this quantity is probably overestimated by the fitting, and some error in the temperature estimate is also expected.  Nevertheless the results suggest that our simulations probe the temperature regime $k_\mathrm{B}T\gtrsim\mu$ in which the thermal friction on the vortex is expected to be appreciable \cite{Fedichev99}, consistent with our observations.
%%%%%%%%%%%%%%%%%%%%%%%%%%%%%%%%%%%%%%%%%%%%%%%%%%%%%%%%%%%
\subsection{Moment of inertia and vortex displacement}
In Fig.~\ref{fig:fns_of_energy}(d) we plot the classical moment of inertia of the noncondensate (crosses), $\langle \Theta_\mathrm{c} \rangle_\mathrm{th}$ (see Sec.~\ref{subsubsec:cloud_rotation}).  We find that this quantity increases steadily with the field energy as atoms are excited out of the condensate into the thermal cloud.  We also plot the (quantum) moment of inertia of the condensate orbital (plusses), which we define as $\Theta_0\equiv\langle L_z \rangle_0/\Omega_\mathrm{c}$, where the angular momentum $\langle L_z\rangle_0$ and rotation frequency $\Omega_\mathrm{c}$ of the condensate are those discussed in the previous sections.  To give an indication of the variation of estimates in this derived quantity, we include the standard deviation estimated by adding the standard deviations of $\langle L_z\rangle_0$ and $\Omega_\mathrm{c}$ in quadrature \cite{noteR}.  We note that the moment of inertia of the condensate decays steadily with increasing $E$, following $\langle L_z \rangle_0$, and by $E=1.15E_\mathrm{g}$ is essentially zero.  Above this energy the condensate's angular momentum is basically zero (see the remarks in Sec.~\ref{subsec:cfrac_and_angmom}), i.e. the condensate fails to respond to the rotation imposed by the thermal cloud, and its moment of inertia is thus zero.  

On this plot we also include the displacement of the vortex from the trap center (circles).  For consistency with the other measurements we calculate this by tracking vortex trajectories over 100 successive intervals of 10 cyc.\ (starting from $t=9000$ cyc.) and averaging the measured radii.  Bars on the marker again indicate standard deviations of the estimates over the 100 periods.  The precession radius increases steadily as the energy is increased.  By $E=1.13E_\mathrm{g}$, the precessing vortex is close to the edge of the condensate, and so the simple vortex tracking algorithm begins to have difficulty tracking this single vortex in proximity to the proliferation of phase defects undergoing constant creation and annihilation in the condensate periphery \cite{Wright08}.  We therefore do not attempt to include data for these higher energies on the plot.  It is clear however that the vortex displacement increases as the condensate angular momentum decreases, as is the case in the zero-temperature GP theory \cite{Butts99,Ballagh99,Vorov05}.  
%%%%%%%%%%%%%%%%%%%%%%%%%%%%%%%%%%%%%%%%%%%%%%%%%%%%%%%%%%%%%%%%%%%%%%%%%%%%%%%%%%%%%%%%%%%%%%%%%%%%%%%%%%%%%%%%%%%%%%%%%%%%%%%%%%%%
\section{Summary and Conclusions}\label{sec:Conclusions}
We have carried out classical-field simulations of a precessing vortex in a finite-temperature Bose-Einstein condensate.  Employing a microcanonical ergodic-evolution approach, we find that randomized states of the classical field subject to an additional angular-momentum constraint settle to configurations each containing a single vortex precessing under the influence of its own induced velocity field, at thermal and rotational equilibrium with the thermal cloud, so that the mutual friction force between the vortex and the cloud vanishes.

As the classical field condenses into a rotationally symmetry-broken state, a naive application of the Penrose-Onsager test of one-body coherence in an inertial frame yields a condensate fragmented over a set of coherent angular-momentum eigenmodes.  Transforming to a rotating frame such that the fragmentation is eliminated we find that the classical field describes a condensate mode with an off-axis vortex coexisting with a core-filling thermal component, and we also observe the Goldstone mode associated with the rotational symmetry breaking.

Due to the ergodic nature of the classical-field evolution, the rotational phase of the vortex diffuses over time.  Thus care must be taken to separate the short-time fluctuation dynamics which define the coherence of the field from the dynamics on much longer time scales, over which ergodic migration of the field configuration occurs.  We found that extracting the condensate by averaging over short times and averaging over successive estimates so obtained allowed us to access the short-time dynamics of interest while exploiting the ergodic nature of the system on longer time scales to better estimate system observables.

We showed that the angular momentum of the condensed mode decreases dramatically as the energy (and thus temperature) are increased at fixed total angular momentum, and correspondingly the vortex precessional radius and frequency increase until the vortex leaves the condensate, and the condensate becomes irrotational.  We proposed a measure of the cloud angular velocity based on the decomposition of the field into condensed and noncondensed components and found qualitative agreement with the vortex precession frequency.

Possible extensions of the work presented here are to nonequilibrium scenarios of vortex decay and to multiple vortex configurations.  We also expect the approaches developed here to be of more general applicability in extracting dynamical characteristics of condensates in ergodic classical field studies.
%%%%%%%%%%%%%%%%%%%%%%%%%%%%%%%%%%%%%%%%%%%%%%%%%%%%%%%%%%%%%%%%%%%%%%%%%%%%%%%%%%%%%%%%%%%%%%%%%%%%%%%%%%%%%%%%%%%%%%%%%%%%%%%%%%%%
\begin{acknowledgments}
We wish to acknowledge discussions with P.~B.~Blakie, C.~W.~Gardiner, D.~A.~W.~Hutchinson, and B.~G.~Wild.  We are particularly grateful to C.~Fox for assistance with mathematical and numerical aspects of this work.
This work was supported by the New Zealand Foundation for Research, Science and Technology under Contract Nos. NERF- UOOX0703 and UOOX0801.
\end{acknowledgments}
%%%%%%%%%%%%%%%%%%%%%%%%%%%%%%%%%%%%%%%%%%%%%%%%%%%%%%%%%%%%%%%%%%%%%%%%%%%%%%%%%%%%%%%%%%%%%%%%%%%%%%%%%%%%%%%%%%%%%%%%%%%%%%%%%%%%
\appendix
%%%%%%%%%%%%%%%%%%%%%%%%%%%%%%%%%%%%%%%%%%%%%%%%%%%%%%%%%%%%%%%%%%%%%%%%%%%%%%%%%%%%%%%%%%%%%%%%%%%%%%%%%%%%%%%%%%%%%%%%%%%%%%%%%%%%
\section{Dependence on frame of projector}\label{app:AppendixA}
\begin{figure}
	\includegraphics[width=0.45\textwidth]{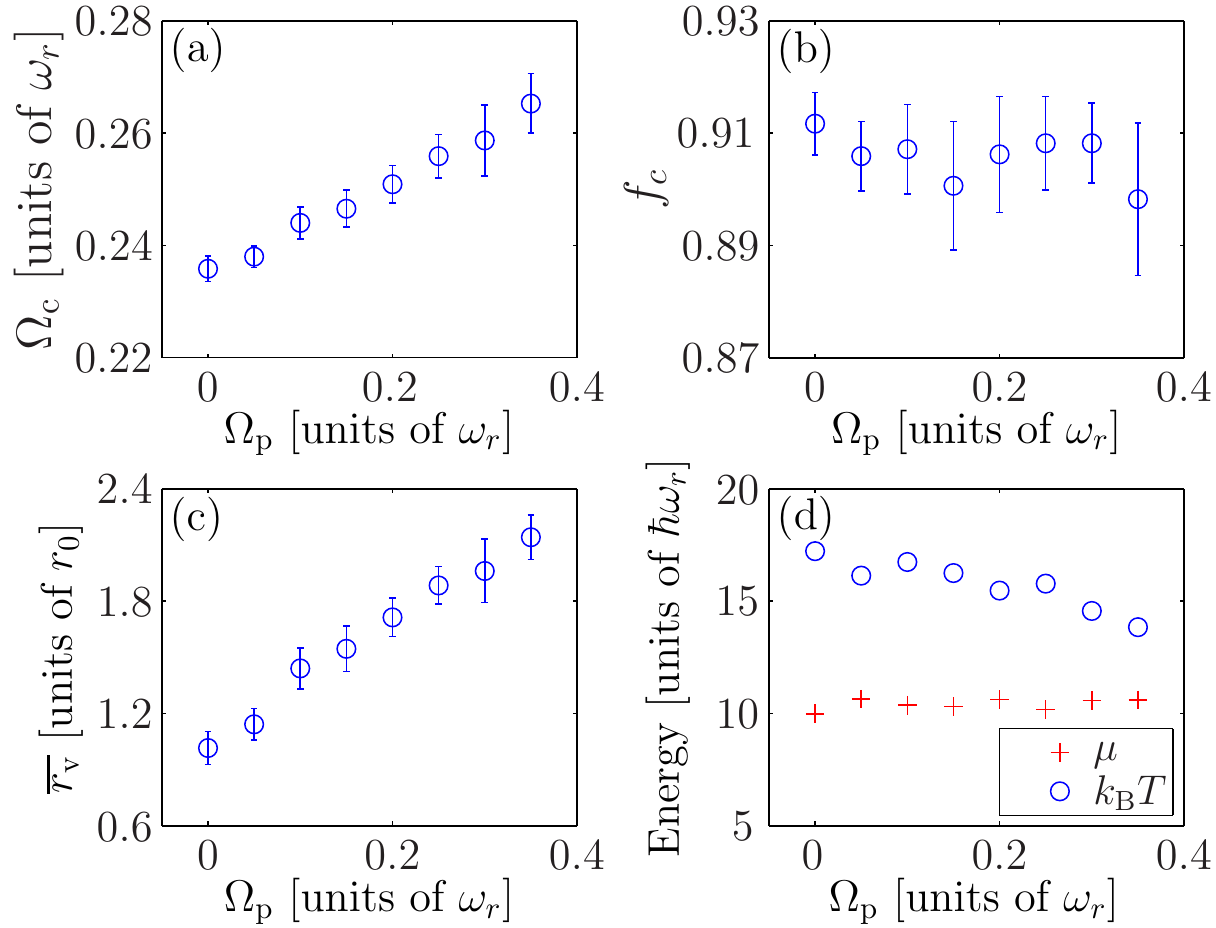}
	\caption{\label{fig:vary_w} (Color online) Dependence of equilibrium system parameters on the frame in which the projection is effected. (a) Angular velocity of the condensate, (b) condensate fraction, (c) vortex precession radius, (d) temperature and chemical potential.}
\end{figure}

%%%%%%%%%%%%%%%%%%%%%%%%%%%%%%%%%%%%%%%%%%%%%%%%%%%%%%%%%%%%%%%%%%%%%%%%%%%%%%%%%%%%%%%%%%%%%%%%%%%%%%%%%%%%%%%%%%%%%%%%%%%%%%%%%%%%
We now consider the effect of the angular velocity $\Omega_\mathrm{p}$ of the frame in which the projector is defined on the equilibrium configurations of the classical-field trajectories.  In irrotational scenarios in which the projector is applied in a laboratory frame, the cutoff energy has a significant effect on the equilibrium attained for given conserved first integrals (energy and normalization).  In the present case, the cutoff is defined by two parameters: the angular velocity of the frame in which the cutoff is effected, and the (rotating-frame) energy at which the cut is made.  As noted in Sec.~\ref{subsubsec:Choice_of_frame}, the (thermodynamic) angular velocity of the classical field at equilibrium is in general not equal to that of the projector.  Here we consider the effect of the choice of projector angular velocity on the classical-field equilibrium.  We vary the projector frequency over the range $0-0.35\omega_r$, and keep the (rotating-frame) cutoff fixed at $E_R = 30\hbar\omega_r$.  There are several consequences of varying the angular velocity $\Omega_\mathrm{p}$.  Firstly the multiplicity $\mathcal{M}$ (i.e., the number of canonical coordinate pairs in the Hamiltonian system) depends strongly on $\Omega_\mathrm{p}$, increasing with increasing $\Omega_\mathrm{p}$.  Moreover the set of modes included varies with $\Omega_\mathrm{p}$, with the condensate band becoming increasingly biased towards single-particle modes with $m>0$ as $\Omega_\mathrm{p}$ is increased \cite{Bradley_PhD}.  In Fig.~\ref{fig:vary_w}(a-d) we present the results of simulations with $E[\psi]=1.10E_\mathrm{g}$, $E_R=30\hbar\omega_r$, $L[\psi]=N_0\hbar$, and $\Omega_\mathrm{p}\in[0,0.35]\omega_r$.  
We find that all these simulations exhibit qualitatively similar equilibrium states, each containing a single precessing vortex, as for the simulation with $E=1.10E_\mathrm{g}$ discussed in the main text.  In Fig.~\ref{fig:vary_w}(a) we plot the angular rotation frequency of the condensate mode, as determined by the procedure of Sec.~\ref{subsubsec:temporal_decoherence}.  The rotation frequency exhibits a clear upward trend as the frame frequency is increased, nevertheless the dependence of the rotation rate on frame frequency is weak, varying by only $\sim0.03\omega_r$ over the range of $\Omega_\mathrm{p}$ considered.  In Fig.~\ref{fig:vary_w}(b) we plot the corresponding condensate fractions.  Again the dependence on the frame frequency is weak; any trend in the condensate fraction is of the same magnitude as the statistical dispersion of the measured values.  In Fig.~\ref{fig:vary_w}(c) we plot the vortex precession radii.  We see that the radius increases steadily, approximately doubling between $\Omega_\mathrm{p}=0$ and $\Omega_\mathrm{p}=0.35\omega_r$.  This increase in precession radius corresponds to a loss of angular momentum (per particle) from the condensate, and is consistent with the observed increase in condensate angular velocity.  In Fig.~\ref{fig:vary_w}(d) we plot the dependence of the chemical potential (crosses) and temperature (circles) on $\Omega_\mathrm{p}$, calculated assuming the condensate angular frequency $\Omega_\mathrm{c}$ as an estimate for the cloud rotation rate.  We observe that the temperature decreases as the frame frequency is increased, consistent with the increase in the number of modes in the condensate band (c.f. \cite{Wright08}).  We conclude then that increasing the frame angular velocity $\Omega_\mathrm{p}$ results in a lowering of the equilibrium temperature, and biases the angular-momentum balance in the system towards the thermal cloud, leading to a higher equilibrium angular velocity.
We emphasize that this dependence is a natural consequence of the projector definition: different values of $\Omega_\mathrm{p}$ define different Hamiltonian systems.  Ultimately, this dependence would be removed by the inclusion of above-cutoff effects \cite{Gardiner03,Davis06,Blakie08}.
%%%%%%%%%%%%%%%%%%%%%%%%%%%%%%%%%%%%%%%%%%%%%%%%%%%%%%%%%%%%%%%%%%%%%%%%%%%%%%%%%%%%%%%%%%%%%%%%%%%%%%%%%%%%%%%%%%%%%%%%%%%%%%%%%%%%
\section{Semiclassical fitting function}\label{app:AppendixB}
%%%%%%%%%%%%%%%%%%%%%%%%%%%%%%%%%%%%%%%%%%%%%%%%%%%%%%%%%%%%%%%%%%%%%%%%%%%%%%%%%%%%%%%%%%%%%%%%%%%%%%%%%%%%%%%%%%%%%%%%%%%%%%%%%%%%
In order to estimate the thermodynamic parameters of the field, we assume an approximate semiclassical description for the thermal atoms.  The description consists of two elements: a classical-field distribution in the phase-space coordinates ($\mathbf{x}$ and $\mathbf{p}$) \cite{Wright08}, and the specification of an appropriate energy cutoff.  In order to construct this distribution, we must take into account two generally disparate rotations: the equilibrium rotation of the thermal atoms (angular velocity $\Omega_\mathrm{th}$) and the rotation of the projector defining the cutoff (angular velocity $\Omega_\mathrm{p}$).
The classical-field distribution takes the general form
\begin{equation}\label{eq:semiclassical_dist1}
	F_\mathrm{c}(\mathbf{x},\mathbf{p}) = \frac{k_\mathrm{B}T}{\epsilon(\mathbf{x},\mathbf{p})-\mu},
\end{equation}
where the semiclassical energy of the thermal atoms (rotating at $\Omega_\mathrm{th}$) is  
\begin{equation}
	\epsilon(\mathbf{x},\mathbf{p}) = \frac{p^2}{2m} + \frac{m\omega_r^2r^2}{2} - \Omega_\mathrm{th}(xp_y-yp_x) + 2\lambda n(r),
\end{equation}
(c.f. \cite{Stringari99}) and where $n(r)$ is the total density of the atomic field, which we will take to be circularly symmetric.  The application of the projector $\mathcal{P}$ defined in a rotating frame enforces a cutoff in the energy, which delimits the accessible region of phase space occupied according to Eq.~(\ref{eq:semiclassical_dist1}).  This region is defined by 
\begin{equation}\label{eq:semiclassical_cutoff_fundamental}
	\frac{p^2}{2m} + \frac{m\omega_r^2r^2}{2} - \Omega_\mathrm{p}(xp_y-yp_x) + 2\lambda n(r) \leq E_R.
\end{equation}
The density of thermal atoms at any spatial position $\mathbf{x}$ is obtained by integrating Eq.~(\ref{eq:semiclassical_dist1})  over the momentum, i.e. $n(\mathbf{x}) = \int_\mathcal{D} d\mathbf{p} F_\mathrm{c}(\mathbf{x},\mathbf{p})$, where the domain $\mathcal{D}$ in momentum space is determined by Eq.~(\ref{eq:semiclassical_cutoff_fundamental}), and is itself a function of the position-space coordinate $\mathbf{x}$.  To perform this integration we make the transformation to the \emph{kinematic} momentum measured in the frame of the cloud $\mathbf{P}=\mathbf{p}-m\bm{\Omega}_\mathrm{th}\times\mathbf{x}$ \cite{Landau60,Stringari99}.  In this representation the integrand becomes circularly symmetric in $\mathbf{P}$, i.e., we now have $F_\mathrm{c}'(\mathbf{x},\mathbf{P})=k_\mathrm{B}T/[\epsilon(\mathbf{x},\mathbf{P})-\mu]$, where 
\begin{equation}
	\epsilon(\mathbf{x},\mathbf{P}) = \frac{P^2}{2m} + \frac{m}{2}(\omega_r^2-\Omega_\mathrm{th}^2)r^2 + 2\lambda n(r).
\end{equation}
In these momentum coordinates the domain $\mathcal{D}$ is circular, however its center is in general displaced from the origin $\mathbf{P}=\mathbf{0}$.  After some simple algebra we find that the domain of integration is
\begin{widetext}
\begin{equation}
	\mathcal{D}(\mathbf{x}): \frac{1}{2m}\Big\{[P_x+m(\Omega_\mathrm{p}-\Omega_\mathrm{th})y]^2 + [P_y-m(\Omega_\mathrm{p}-\Omega_\mathrm{th})x]^2\Big\} \leq E_R - \frac{m}{2}m(\omega_r^2-\Omega_\mathrm{p}^2)r^2 -2\lambda n(r).
\end{equation}
\end{widetext}
As the integrand is circularly symmetric in $\mathbf{P}$, the result of the integral depends only on the radius $r$ of the chosen point in position space, as azimuthal rotations of the coordinate $\mathbf{x}$ simply induce rotations of $\mathcal{D}(\mathbf{x})$ about the origin of momentum space.  We therefore consider without loss of generality $(x,y)=(0,-r)$.  The domain $\mathcal{D}$ thus becomes a circle centered on $P_x=m(\Omega_\mathrm{p}-\Omega_\mathrm{th})r\equiv a$, of radius $b\equiv \sqrt{2m[E_R - (m/2)(\omega_r^2-\Omega_\mathrm{p}^2)r^2-2\lambda n(r)]}$.  Depending on the value of $r$, we may have $a\leq b$ or $b>a$. We adopt polar coordinates $(P,\phi)$, and find for the angle $\Delta\phi$ subtended by $\mathcal{D}$ at a given radius $P$ 
\begin{equation}
	\Delta \phi(P) = \left\{
	\begin{array}{cl}
		0 & P<a-b\;\;(\mathrm{when}\;a>b)\\
        2\pi & P<b-a\;\;(\mathrm{when}\;a<b) \\
        2\cos^{-1}\Big(\frac{P^2-b^2+a^2}{2aP}\Big)  & |a-b|<P<a+b \\
		0 & P>a+b
    \end{array} \right. .
\end{equation}
Integrating in polar coordinates we obtain finally
\begin{eqnarray}
	n(r) &=& \frac{2\pi mk_\mathrm{B}T}{h^2} \int_0^\infty \frac{PdP\Delta\phi(r)}{\frac{P^2}{2m}+\frac{m\omega_r^2r^2}{2} + 2\lambda n(r) - \mu}  \nonumber \\ 
	&=& \frac{1}{\lambda_\mathrm{dB}^2}\Big[I_1(r;\Omega_\mathrm{th},\Omega_\mathrm{p}) + I_2(r;\Omega_\mathrm{th},\Omega_\mathrm{p})\Big],
\end{eqnarray}
where
\begin{equation}
	I_1(r) = \Theta\Big(b - a\Big)\ln\Bigg[\frac{\frac{(b-a)^2}{2m} + \frac{m}{2}(\omega_r^2-\Omega_\mathrm{th}^2)r^2+2\lambda n(r) - \mu}{\frac{m}{2}(\omega_r^2-\Omega_\mathrm{th}^2)r^2+2\lambda n(r)-\mu}\Bigg],  
\end{equation}
and 
\begin{equation}
	I_2(r) = \frac{1}{\pi}\int_{|b-a|}^{a+b}\frac{PdP \cos^{-1}\Big(\frac{\frac{p^2}{2m}+\frac{m}{2}[\omega_r^2-\Omega_\mathrm{p}^2+(\Omega_\mathrm{p}-\Omega_\mathrm{th})^2]r^2-E_R}{(\Omega_\mathrm{p}-\Omega_\mathrm{th})rP}\Big)}{\frac{p^2}{2m}+\frac{m}{2}(\omega_r^2-\Omega_\mathrm{th}^2)r^2-\mu},
\end{equation}
and we have introduced the de Broglie wavelength $\lambda_\mathrm{dB} \equiv \sqrt{2\pi\hbar^2/mk_\mathrm{B}T}$ \cite{Pethick02}.  The integral $I_2$ must be evaluated numerically.  In the limit $\Omega_\mathrm{th}\rightarrow\Omega_\mathrm{p}$ the integral $I_2$ vanishes, and we regain the form
\begin{equation}
	n(r) = \frac{1}{\lambda_\mathrm{dB}^2} \ln\Bigg[\frac{E_R-\mu}{(\omega_r^2-\Omega_\mathrm{th}^2)r^2+2\lambda n(r) - \mu}\Bigg],
\end{equation} 
employed in \cite{Wright08}.

\bibliographystyle{prsty}

\begin{thebibliography}{100}
\bibitem{noteA}
{The projected field $\psi(\mathbf{x})$ is thus not strictly a field in the
  traditional sense \cite{Sudarshan74}, however we will retain the usage of the
  term in this work.}
\bibitem{noteB}
{We note that geometry and projector definition of \cite{Blakie05} yield a
  conserved angular momentum about the $x$~axis, however any nonzero angular
  momenta in the fields evolved in that work were incidental, and of
  essentially negligible magnitude, as noted by the authors of that paper.}
\bibitem{noteV}
{The appearance of the inertial correction $(\Delta H)_\mathrm{inertia} \equiv
  -\Omega_\mathrm{p}L_z$ to the single particle Hamiltonian \cite{Landau60}
  serves only to ensure that the rotating-frame equations of motion faithfully
  describe the same dynamics as the fundamental inertial-frame equations.}
\bibitem{noteD}
{In practice we average momentum-space densities taken over consecutive 10
  cycle periods. Aside from fluctuations close to the origin $\mathbf{k}=0$
  arising from the vortex precession, we observe the densities to settle to
  their (fluctuating) values well before $t=9000$ cyc.}
\bibitem{noteE}
{We use the peak density of our principal ground state to estimate the healing
  length $\eta$.}
\bibitem{noteF}
{The cartesian-grid vortex-tracking algorithm employed here produces a
  systematic bias in the vortex position estimate on a sub-grid-point scale
  (much less than $r_0$). We have subtracted this dc component of the series
  $\{z_\mathrm{v}\}$ before calculating the power spectrum presented in
  Fig.~\ref{fig:equil_vortex_motion}(b).}
\bibitem{noteG}
{This property of the many-body system persists in the classical-field model
  \cite{Blakie07a}, but note however the effect of the projector on large
  dipole oscillations \cite{Bradley05}.}
\bibitem{noteH}
{Vortices may of course occur in all such simulations, we use the phrase
  vortex-free here to refer to simulations in which any appearance of vortices
  is incidental, and sufficiently transient as to not affect the statistics
  under consideration significantly.}
\bibitem{noteI}
{We adopt here the convention for Bogoliubov-mode component labelling of
  \cite{Castin98, Morgan00}.}
\bibitem{noteJ}
{We note that the next most highly occupied eigenmodes, $\chi_3(\mathbf{x})$
  and $\chi_4(\mathbf{x})$, have the form of vortex modes with charges $q=+3$
  and $q=-1$ respectively. These circulations correspond respectively to those
  of the $u$ and $v^*$ components of the linear excitation of the GP vortex
  with angular momentum $m=+2$ relative to the GP state \cite{Mizushima04}.}
\bibitem{noteK}
{We note that the smaller frequency peaks in the spectrum of each coefficient
  correspond to the quasi-uniform phase rotation of other angular-momentum
  components. This arises due to the imperfect nature of the PO decomposition
  resulting from (e.g.) the finite sample size used to construct $\rho_{ij}$
  \cite{Blakie05}.}
\bibitem{noteU}
{Note that the Goldstone mode only happens to be the second most highly
  occupied mode in the present case as its spurious `population' is larger than
  that of all other noncondensate modes.}
\bibitem{noteL}
{It could indeed be said that this is the definitive distinction of
  interpretation between the time-dependent GP theory and the time-dependent
  classical-field theory.}
\bibitem{noteN}
{As we vary $\Omega_\mathrm{m}$ in increments of $\delta \Omega =
  10^{-4}\omega_r$, smaller than the frequency resolution afforded by the
  sampling period ($\Delta\Omega = 1/T_\mathrm{samp}=2\pi\times0.01\omega_r$),
  we resolve the profile $n_0(\Omega_\mathrm{m})
  \sim\mathrm{sinc}^2[(\Omega_\mathrm{m}-\Omega_\mathrm{m}^0)/\Delta\Omega]$ of
  the peak of $n_0(\Omega_\mathrm{m})$ [inset to
  Fig.~\ref{fig:fracs_and_occs_corot}(b)] resulting from its convolution with
  the boxcar function representing the finite sample time $T_\mathrm{samp}$
  \cite{Press92}. The location of the central peak of this form provides the
  best estimate for $\Omega_\mathrm{c}$ in an effective least-squares sense
  \cite{noteO}.}
\bibitem{noteW}
{In the short-time averages (Sec.~\ref{subsec:PO_rotating_frames}) the
  Goldstone mode is typically not differentiated from the proper noncondensed
  modes, as its `occupation' is likely to be small. We therefore simply regard
  all modes orthogonal to the condensate mode $\chi_0(\mathbf{x})$ as
  comprising the noncondensate field here.}
\bibitem{noteT}
{Where the condensate is vortex-free, we expect that the calculated condensate
  fraction should be independent of the frame in which the covariance matrix is
  constructed. However the condensate fraction often exhibits a small rise
  around $\Omega_\mathrm{m}=1\omega_r$, which we associate with a persistent
  dipole excitation of the field.}
\bibitem{noteR}
{The variations in these quantities are of course not independent, however the
  contribution of the variation in estimates of $\langle L_z\rangle_0$
  dominates to such an extent that this is of no concern for the indicative
  representation we seek here.}
\end{thebibliography}

%%%%%%%%%%%%%%%%%%%%%%%%%%%%%%%%%%%%%%%%%%%%%%%%%%%%%%%%%%%%%%%%%%%%%%%%%%%%%%%%%%%%%%%%%%%%%%%%%%%%%%%%%%%%%%%%%%%%%%%%%%%%%%%%%%%%
\end{document}